%% file: main.tex
\documentclass[sigplan,nonacm]{acmart}

\input{packages}

\usepackage{graphicx}
\usepackage{subcaption}
\usepackage[]{hyperref}

\newcommand{\circled}[1]{%
    \tikz[baseline=(char.base)]{
        \node[shape=circle,draw,inner sep=0.5pt,fill=black,text=white,minimum size=1em] (char) {#1};
    }%
}

\newcommand{\SysName}{\textup{LoongTrain}\xspace}

\begin{document}

\title{\SysName: Efficient Training of Long-Sequence LLMs with Head-Context Parallelism}

\author{Diandian Gu}
\affiliation{%
  \institution{School of Computer Science \\ Peking University}
  \country{}
}

\author{Peng Sun}
\affiliation{%
  \institution{Sensetime Research \& \\ Shanghai AI Laboratory}
  \country{}
}

\author{Qinghao Hu}
\affiliation{%
  \institution{S-Lab, NTU \& \\ Shanghai AI Laboratory}
  \country{}
}

\author{Ting Huang}
\affiliation{%
  \institution{Sensetime Research}
  \country{}
}

\author{Xun Chen}
\affiliation{%
  \institution{Sensetime Research}
  \country{}
}

\author{Yingtong Xiong}
\affiliation{%
  \institution{Shanghai AI Laboratory}
  \country{}
}

\author{Guoteng Wang}
\affiliation{%
  \institution{Shanghai AI Laboratory}
  \country{}
}

\author{Qiaoling Chen}
\affiliation{%
  \institution{S-Lab, NTU}
  \country{}
}

\author{Shangchun Zhao}
\affiliation{%
  \institution{Tencent}
  \country{}
}

\author{Jiarui Fang}
\affiliation{%
  \institution{Tencent}
  \country{}
}

\author{Yonggang Wen}
\affiliation{%
  \institution{Nanyang Technological University}
  \country{}
}

\author{Tianwei Zhang}
\affiliation{%
  \institution{Nanyang Technological University}
  \country{}
}

\author{Xin Jin}
\affiliation{%
  \institution{School of Computer Science \\ Peking University}
  \country{}
}

\author{Xuanzhe Liu}
\affiliation{%
  \institution{School of Computer Science \\ Peking University}
  \country{}
}


\input{0_Abstract}

 \maketitle
\pagestyle{plain}
\input{1_Introduction}

\input{2_Background}
\input{2_1_Motivation}

\input{3_SystemDesign}

\input{4_Evaluation}

\input{6_Conclusion}

\bibliographystyle{plain}%acm
\bibliography{references.bib}

% \newpage
\input{7_Appendix}

\newpage

\end{document}

%% file: packages.tex
\usepackage{xspace}
\usepackage{mathtools} % upgrade from amsmath
\usepackage{bm}
\usepackage{enumitem}
\usepackage{textcomp}
\usepackage{xcolor}
\usepackage{changepage}
\usepackage{framed}

\usepackage{hyperref} 
\usepackage{graphicx}
\usepackage{subcaption}
\usepackage{tikz}

\usepackage{tabularx}
\usepackage{multirow}
\usepackage{booktabs}
\usepackage{makecell}
\usepackage[figuresleft]{rotating}
\usepackage[flushleft]{threeparttable}

\usepackage{algorithmicx}
\usepackage{algorithm}
\usepackage[noend]{algpseudocode}
\usepackage{amsmath}
\usepackage{xcolor}
\usepackage{colortbl}

\algnewcommand\algorithmicinput{\textbf{Input:}}
\algnewcommand\Input{\item[\algorithmicinput]}
\algnewcommand\algorithmicoutput{\textbf{Output:}}
\algnewcommand\Output{\item[\algorithmicoutput]}

\algnewcommand{\LeftComment}[1]{\Statex \(\triangleright\) #1}

%% file: 0_Abstract.tex
\begin{abstract}
Efficiently training LLMs with long sequences is important yet challenged by the massive computation and memory requirements. Sequence parallelism has been proposed to tackle these problems, but existing methods suffer from scalability or efficiency issues. We propose \SysName, a novel system to efficiently train LLMs with long sequences at scale. The core of \SysName is the 2D-Attention mechanism, which combines both \textit{head-parallel} and \textit{context-parallel} techniques to break the scalability constraints while maintaining efficiency. We introduce Double-Ring-Attention and analyze the performance of device placement strategies to further speed up training. We implement \SysName with the hybrid ZeRO and \textit{Selective Checkpoint++} techniques.
Experiment results show that \SysName outperforms state-of-the-art baselines, i.e., DeepSpeed-Ulysses and Megatron Context Parallelism, in both end-to-end training speed and scalability, and improves Model FLOPs Utilization (MFU) by up to 2.88$\times$.
\end{abstract}

%% file: 1_Introduction.tex
\section{Introduction}
With the emergence of Large Language Models (LLM) in recent years, researchers have investigated and proposed many advanced training methodologies in a distributed way, such as data parallelism (DP)~\cite{KrizhevskySH12, paszke2019pytorch, li2014scaling, li2014communication}, tensor parallelism (TP)~\cite{DeanCMCDLMRSTYN12},  pipeline parallelism (PP)~\cite{GPipe, AthlurSSRK22}, PyTorch FSDP~\cite{PyTorchFSDP}, and automatic parallelization frameworks~\cite{Alpa}. 
Recently, LLMs with long sequences have driven the development of novel applications that are essential in our daily lives, including generative AI~\cite{ni2023recent} and long-context understanding~\cite{beltagy2020longformer, zhou2021document,ding2023longnet}. With the increased popularity of ChatGPT, long dialogue processing tasks have become more important for chatbot applications than ever~\cite{touvron2023llama}.
In addition to these scenarios for language processing, Transformer-based giant models also achieve impressive performance in computer vision~\cite{zhang2020span, arnab2021vivit, yuan2021tokens} and AI for science~\cite{bi2023accurate, ai4science}, where inputs with long sequences are critical for complex tasks such as video stream processing~\cite{ruan2022survey} and protein property prediction~\cite{chandra2023transformer}.

Training LLMs with long sequences requires massive memory resources and computation. To tackle these challenges, sequence parallelism (SP) has been proposed~\cite{DeepspeedUlysses, lightseq,BPT2, megatroncp}, which can be basically divided into two categories: head parallelism (HP) \cite{DeepspeedUlysses} and context parallelism (CP)  \cite{BPT2, megatroncp}. %\todo{\qh{Maybe need some brief introduction (one sentence for each one).}}
In Attention blocks, HP methods keep the whole sequence and compute attention
for different heads in parallel, while CP methods split the QKV (Query, Key, and
Value) tensors into chunks along the sequence dimension. 
%However, they suffer from two limitations. 
% However, neither of them is suitable for training long-sequence LLMs on a large scale.
However, both face limitations when applied to extremely-long-sequence LLMs at a large scale.
\textbf{First, HP meets the scalability issue.} In HP, the degree of SP inherently cannot exceed the number of attention heads \cite{DeepspeedUlysses}.
% \textbf{First, when the sequence is long, HP approaches cannot scale out the number of GPUs used.}
%The number of GPUs in training is strictly constrained the number of attention heads. 
%The degree of DP and PP are also restricted by the global batch size.\todo{\qh{Why mention DP\&PP here? I think mention GQA would be better}} 
Therefore, there is an upper bound for the degree that HP can scale out. 
%Even though the degree of data parallelism (DP) and pipeline parallelism (PP) cannot scale out to a very large scale because of the restriction of global batch size.
\textbf{Second, CP meets the communication inefficiency issue.} 
% \textbf{Second, CP approaches, such as Ring-Attention \cite{lightseq} and Megatron Context Parallelism \cite{megatroncp}, suffer from  inefficiencies due to high communication overhead.} 
CP \cite{BPT2, megatroncp} employs a peer-to-peer (\texttt{P2P}) communication primitive. However, \texttt{P2P} encounters issues of low intra-node bandwidth utilization and low inter-node network resource utilization. This bottleneck makes it challenging to overlap communication with computation when scaling out the context-parallel dimension. For example, our experiments show that Ring-Attention can spend 1.8$\times$ time on communication than on computation when running Grouped Query Attention (GQA) on 64 GPUs with a sequence length of 128K (Figure~\ref{fig:motivation-attention-time}(d)).

% Second, CP approaches (such as Ring-Attention \cite{lightseq} and Megatron Context Parallelism \cite{megatroncp}) are inefficient due to the high communication overhead. These approaches introduce a peer-to-peer (\texttt{P2P}) communication bottleneck, which 
% is caused by low intra-node bandwidth utilization and low-inter-node NIC utilization.
% It is difficult to overlap communication with computation when scaling out the Context-Parallel dimension.
\begin{table}%[htbp]
\centering
\small % Use smaller font size
\begin{tabular}{@{}l@{\hspace{0.15cm}}l@{\hspace{0.15cm}}|@{\hspace{0.15cm}}l@{\hspace{0.15cm}}l@{}}
\toprule
$S$   & Sequence Length (Tokens) & $d_{sp}$ & Sequence Parallel Size \\
$H$   & Number of Attention Heads & $d_{dp}$ & Data Parallel Size \\
$H_{kv}$   & Number of KV Heads & $d_{hp}$ &  Head Parallel Size \\
$D$   & Hidden Dimension Size & $d_{cp}$ &  Context Parallel Size \\
$B$   & Global-Batch Size (Tokens) &  $w$ &  Inner Ring Size \\
% $N$ & Number of GPUs  &  $w$ &  Inner Ring Size \\ 
\bottomrule
\end{tabular}
\caption{Notations used in this paper.}
\label{tab:transformer-parameters}
\end{table}

To bridge these gaps, we propose \SysName, an effective training framework for long-sequence LLMs on large-scale GPU clusters. Our key idea is to address the scalability constraints of HP while mitigating the inefficiencies of CP by introducing a novel \textit{2D-Attention} mechanism. This mechanism parallelizes attention across both HP and CP dimensions. Specifically, it distributes the QKV tensors across GPUs based on the head dimension and partitions these tensors into chunks within the CP dimension. 
By doing so, \SysName enhances scalability through the integration of CP and reduces the number of \texttt{P2P} steps by confining the CP dimension size. In addition, this design provides more opportunities for computation-communication overlap.

To further improve the communication efficiency of Attention blocks in certain circumstances, we introduce \textit{Double-Ring-Attention}, which utilizes all of the inter-node NICs efficiently for higher peer-to-peer communication bandwidth. We also analyze how different \textit{placement strategies} can boost the communication efficiency in different 2D-Attention configurations. Finally, we implement advanced techniques such as applying ZeRO across both DP and PP dimensions and a whitelist-based gradient checkpointing mechanism \textit{Selective Checkpoint++} to further improve the end-to-end LLM training performance.  Evaluation results on training LLMs with up to 1M sequences show that \SysName can bring up to 2.88$\times$ performance improvement compared to existing state-of-the-art solutions.

\SysName has been deployed to train multiple long-sequence LLMs within our organization. The system is implemented within our internal training framework, which can be accessed at \url{https://github.com/InternLM/InternEvo}.

% Our contributions are as follows.
% \begin{itemize}[topsep=0pt,parsep=0pt]
% \item We design 2D-Attention to break the scalability limits of training LLMs with long sequences while maintaining high efficiency. %2D-Attention
% \item We propose a novel Double-Ring-Attention mechanism and analyze the performance of different device placement strategies, to gain additional performance improvement in Attention blocks. 
% \item We implement a system prototype of \SysName with advanced techniques (hybrid ZeRO and Selective Checkpoint++) to speed up the end-to-end training.
% \item Evaluation results on training LLMs with up to 1M sequences show that \SysName can bring up to 2.54$\times$ Tokens per GPU per Second (TGS) improvement compared to existing state-of-the-art solutions.
% \end{itemize}

%% file: 2_Background.tex
\section{Background}

\subsection{LLM Architecture with MHA/GQA}

\begin{figure}[htbp]
    \centering
    \includegraphics[width=0.465\textwidth]{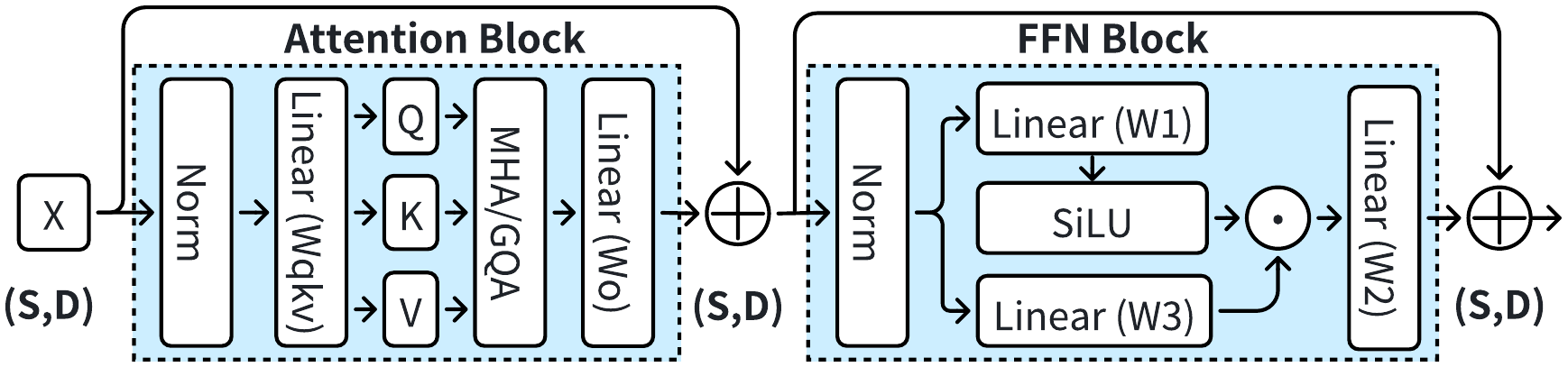}
    \caption{A typical Transformer layer contains an Attention block and a Feed-Forward Network (FFN) block.}
    \label{fig:transformer_arch}
\end{figure}

LLMs like GPT \cite{GPT3} and LLaMA \cite{LLaMA} utilize the Transformer architecture \cite{Attention}, which consists of multiple layers. As shown in Figure \ref{fig:transformer_arch}, each layer includes an Attention block and a Feed-Forward Network (FFN) block. Within the Attention block, a linear module projects the input tensor into three tensors: Query (\( Q \)), Key (\( K \)), and Value (\( V \)), which are used for attention computation. Then, each layer includes an FFN, which operates independently on each position within the sequence.
\( \text{FFN}(x) = W_2(\text{SiLU}(W_1(x)) \times W_3(x)) \), where $W_1, W_2, W_3$ are all linear modules.

Multi-Head Attention (MHA) \cite{MHA} splits \( Q \), \( K \), and \( V \) into \( H \) heads. Suppose the original \( Q \), \( K \), and \( V \) tensors have the shape \( (S, D) \). They will be reshaped to \( (H, S, D/H) \). MHA performs attention computation for each head independently and then combines the outputs of all heads. 
Grouped Query Attention (GQA) \cite{GQA} divides the \( H \) query heads into \( G \) groups, with each group sharing a single set of KV heads. In this case, the transformed \( K \) and \( V \) tensors have \( H_{kv} = H / G \), resulting in a shape of \( (H_{kv}, S, D/H) \). For example, LLaMA3-8B \cite{llama3modelcard} employs GQA with \( H_{kv} = 8 \) and \( H = 32 \) .

\subsection{Distributed LLM Training}
Hybrid parallelism \cite{Megatron-LM} and Zero Redundancy Optimizer (ZeRO) \cite{ZeRO} are commonly employed to train LLMs at scale. Specifically, data parallelism (DP) divides input data into chunks, distributing them across multiple GPUs to parallelize training. Tensor parallelism (TP) distributes model parameters across GPUs along specific dimensions, enabling parallel computation of the model layers \cite{TP}. Pipeline parallelism (PP)  splits layers of a model into multiple stages, distributing them across GPUs \cite{GPipe,pipedream}. 
Each pipeline stage depends on the outputs of previous stages, leading to computation stalls known as pipeline bubbles. Advanced pipeline schedulers, such as 1F1B \cite{pipedream} and ZeRO-Bubble \cite{zerobubble}, have been proposed to reduce the bubble ratio.
ZeRO \cite{ZeRO} addresses redundant memory usage across DP ranks. 
ZeRO-1 partitions optimizer states across GPUs, ensuring each GPU stores only a fraction of the optimizer state. ZeRO-2 extends this by also sharding gradients, and ZeRO-3 further distributes model parameters.

% \subsection{Sequence Parallelism}

\begin{figure}%[htbp]
    \centering
    \includegraphics[width=0.465\textwidth]{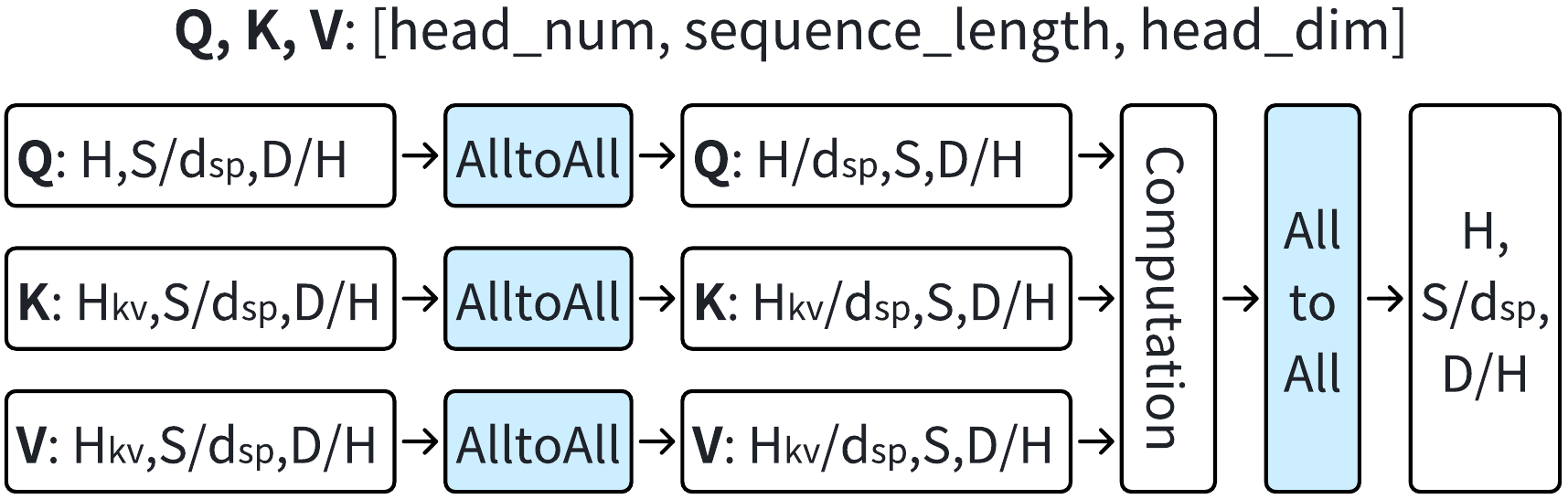}
    \caption{Ulyssess-Attention performs head-parallel computation across GPUs with two steps of \texttt{AlltoAll}.}
    \label{fig:ulyssess-attentions}
\end{figure}

To support long-sequence training, sequence parallelism (SP) has emerged as an effective technique to mitigate activation memory footprints  \cite{DeepspeedUlysses,Nvidia3,lightseq}. In SP, the input and output tensors of each Transformer layer are partitioned into $d_{sp}$ chunks along the sequence dimension. 
Megatron-LM integrates SP with TP across different modules \cite{Nvidia3}. 
Specifically, TP is utilized to parallelize the linear modules, while SP is applied to normalization and dropout modules. 
To ensure consistency in computational results, Megatron-LM incorporates necessary \texttt{AllGather} and \texttt{ReduceScatter} operations to transfer activations during training. However, as the sequence length increases, the communication overhead associated with transferring activations also grows, leading to significant communication challenges \cite{DeepspeedUlysses,hu2024characterization}. 

To address this problem in the integration of SP and TP, recent approaches implement SP across all linear modules and utilize ZeRO-3 to reduce memory footprints. This eliminates the need for collective communications on activations. They perform \texttt{AllGather}  to collect the parameters of linear modules before computation, which do not increase with the sequence length. Following this strategy, two methods have been introduced to facilitate distributed attention computation: Ulysses-Attention \cite{DeepspeedUlysses}  and Ring-Attention \cite{lightseq,BPT2}, as described below. 

\subsection{Distributed Attention}

Ulysses-Attention \cite{DeepspeedUlysses} performs head-parallel computation across GPUs (\(d_{hp} = d_{sp}\)), as depicted in Figure \ref{fig:ulyssess-attentions}. Given the QKV tensors, which are split along the sequence dimension, Ulysses-Attention first performs \texttt{AlltoAll} to ensure that each GPU receives the complete sequence of QKV for \(H/d_{sp}\) heads. Each GPU then computes the attention for different heads in parallel. Finally, another \texttt{AlltoAll} operation gathers the results across the head dimension while re-partitioning along the sequence dimension.

\begin{figure}%[htbp]
    \centering
    \begin{subfigure}[b]{0.235\textwidth}
        \centering
        \includegraphics[width=\textwidth]{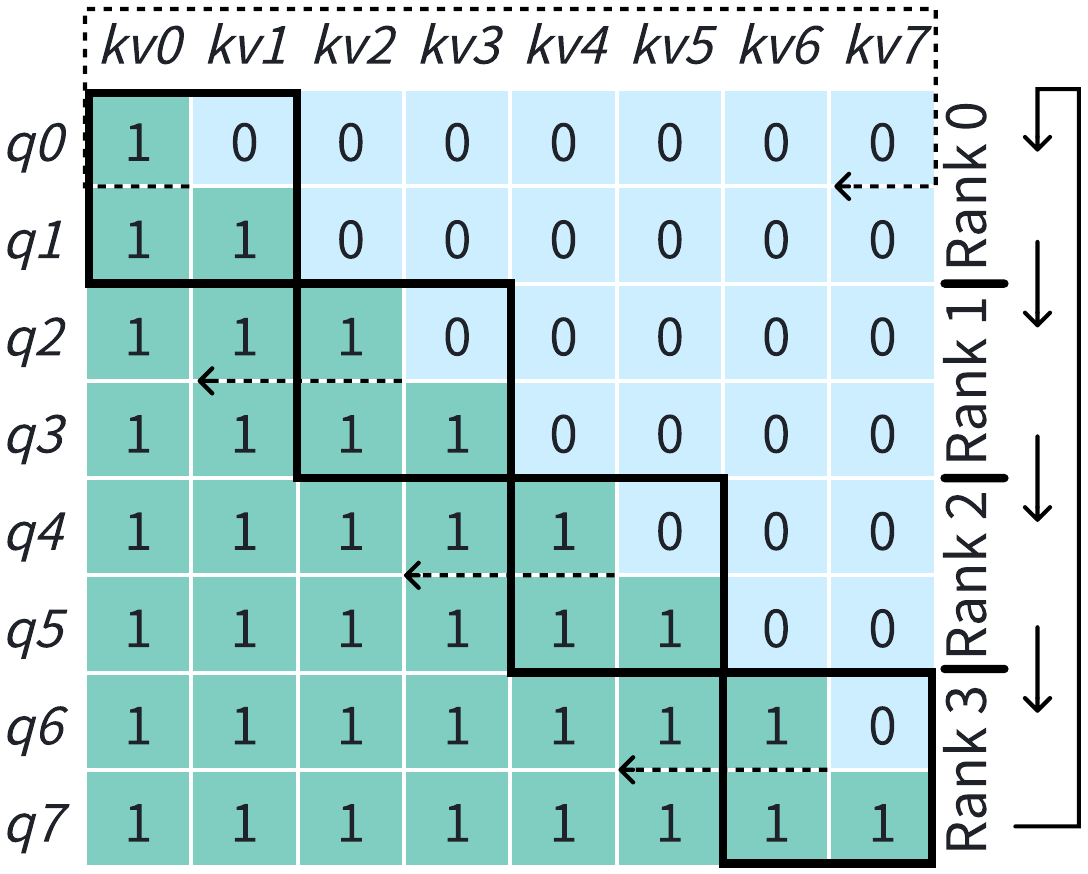}
        \caption{Without Load Balance}
        \label{fig:ring-attention}
    \end{subfigure}
    \begin{subfigure}[b]{0.235\textwidth}
        \centering
        \includegraphics[width=\textwidth]{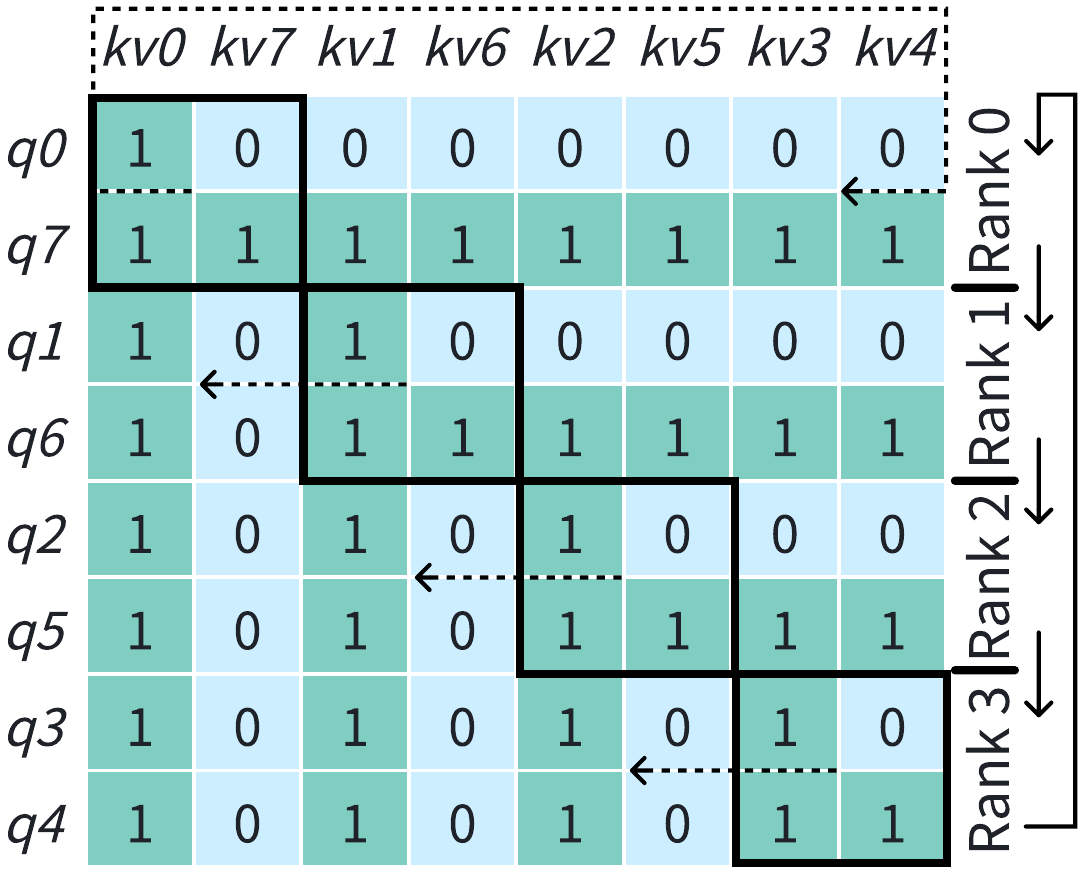}
        \caption{With Load-Balance}
        \label{fig:load-balance-ring-attention}
    \end{subfigure}
    \caption{Ring-Attention performs context-parallel computation, and organizes communication in a ring fashion. 1 or 0 represents that whether there is computation between QKV.}
    \label{fig:two-ring-attentions}
\end{figure}

Ring-Attention \cite{lightseq,BPT2} leverages blockwise attention  \cite{self-attnnotneedon2memory,BPT1,flashattn1} and performs context-parallel computation (\(d_{cp} = d_{sp}\)), as shown in Figure \ref{fig:two-ring-attentions}. This method partitions QKV tensors into chunks along the sequence dimension, with each GPU initially assigned one chunk. For each query chunk, its corresponding attention output is computed by iterating over all KV chunks. Communication is organized in a ring fashion, where each GPU simultaneously sends and receives KV chunks, allowing communication to be overlapped with computation. FlashAttention \cite{flashattn1} can still be used to maintain the IO-aware benefits of memory-efficient computation.
However, the standard Ring-Attention approach is not load-balanced when applying a causal attention mask, since only the lower triangular portion of the matrix needs to be computed. To address this issue, several methods have been proposed, such as DistFlashAttn \cite{lightseq} and Striped-Attention \cite{StripedAttention}. As shown in Figure \ref{fig:two-ring-attentions}(b), Megatron-LM reorders the input sequence tokens along the sequence dimension to achieve load balance in its implementation. In this paper, Ring-Attention is assumed to be load-balanced by default.

%% file: 2_1_Motivation.tex
\section{Motivation \& Observation}

% With existing memory-efficient training mechanisms, a single GPU can train models with sequences containing 256K tokens, and four GPUs can handle sequences with 1M tokens. 
Given the long computation time of LLM training, especially with long sequences, it is essential to scale long-sequence model training to large-scale clusters. %with thousands of GPUs. 
However, current SP approaches face two significant challenges: limited scalability and high communication overhead.

% This scenario introduces two challenges: the scalability of current parallelism methods and the communication overhead introduced by these methods.
\begin{figure}%[htbp]
    \centering
    \begin{subfigure}[b]{0.235\textwidth}
        \centering
        \includegraphics[width=\textwidth]{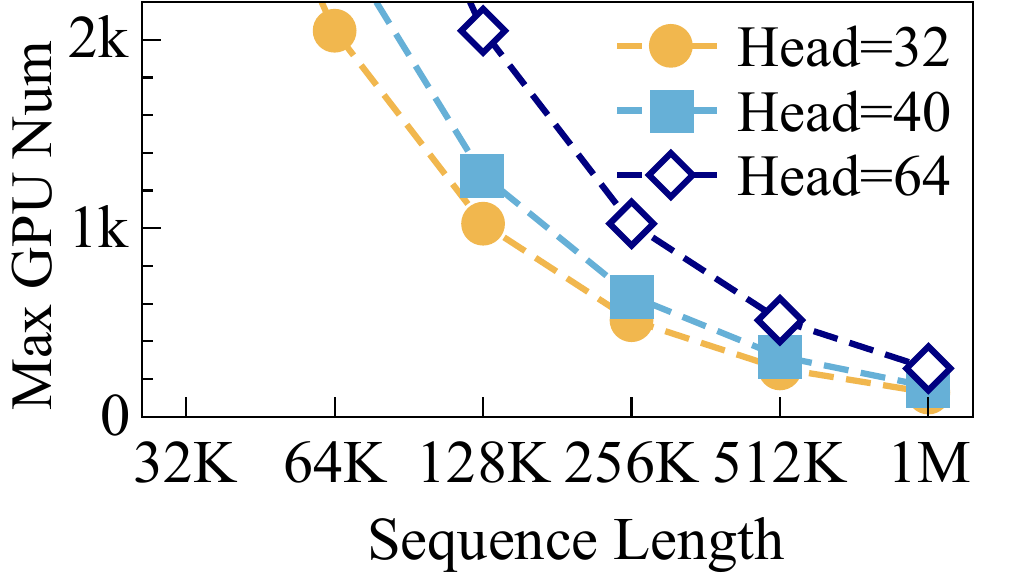}
        \caption{Maximum GPU Scalability}
        \label{fig:scalibility_problem_nopp}
    \end{subfigure}
    \begin{subfigure}[b]{0.235\textwidth}
        \centering
        \includegraphics[width=\textwidth]{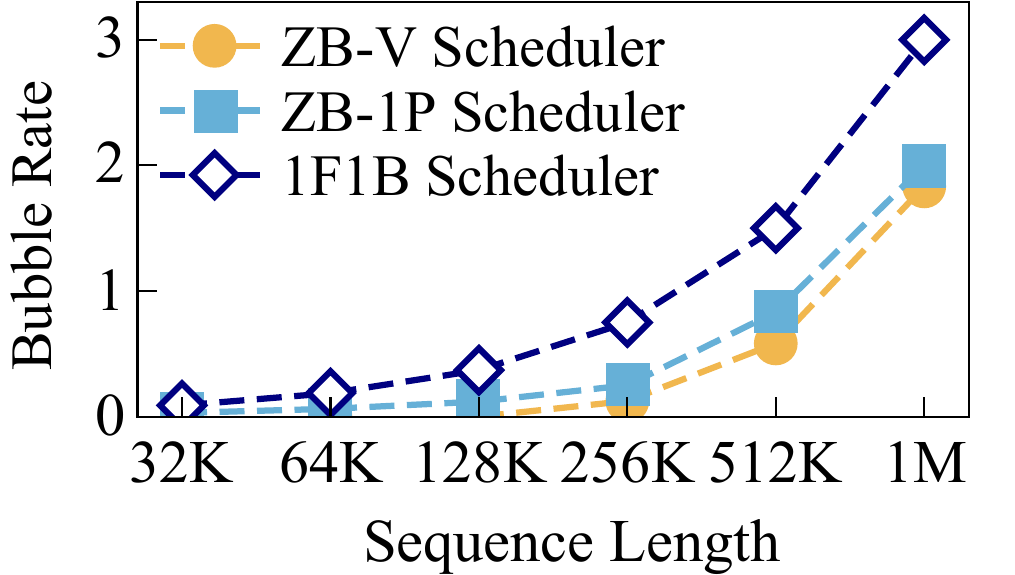}
        \caption{Pipeline Bubble Rate}
        \label{fig:scalibility_problem_with_pp}
    \end{subfigure}
    \caption{Limited scalability of Ulysses-Attention constrained by a global batch size of 4M tokens. (a) Maximum GPU scalability without Pipeline Parallelism. (b) Pipeline bubble rate, using $d_{dp}=4, d_{sp}=64, d_{pp}=4$ on 1024 GPUs.}
    \label{fig:scalibility_problem}
\end{figure}
\subsection{Limited Scalability of Ulysses-Attention}

Ulysses-Attention cannot scale long-sequence training to large-scale clusters due to the limitations in the maximum degrees of SP, DP, and PP.
First, SP is sensitive to the number of attention heads. 
When using MHA, the SP degree cannot exceed the number of attention heads; while in the case of GQA, the SP degree is limited by the number of key/value heads. For instance, LLaMA3-8B uses GQA with 8 key/value heads, meaning that the maximum SP degree is 8 when using Ulysses-Attention. Even if we repeat key/value heads, as detailed in Section \ref{subsec: 2d_overview}, the maximum SP degree remains 32.

It is impractical to rely on increasing the degree of DP to scale out the training process due to the constraint of the global batch size.  For instance, when training a Transformer model with 32 attention heads and employing a global batch size of 4M tokens—as exemplified in the world model training \cite{liu2024world}—and a sequence length of 1M tokens, the maximum attainable degree of DP is 4. Under these conditions, the training process can only be scaled up to 128 GPUs when utilizing Ulysses-Attention. The maximum number of GPUs that Ulysses-Attention could use within the constraint of a 4M global batch size is illustrated in Figure \ref{fig:scalibility_problem} (a).

While we can scale out long-sequence training to more GPUs by increasing the degree of PP, it can lead to a high bubble rate. Due to the global batch size constraint, we have a limited number of micro-batches, which introduce a significant bubble rate. As shown in Figure \ref{fig:scalibility_problem}(b), the bubble rate reaches 2 even under zero-bubble mechanisms, such as the ZB-V and ZB-1P schedulers \cite{zerobubble}. This level of inefficiency is unacceptable for effective LLM training.

\subsection{Inefficient Performance of Ring-Attention}
\label{sec:motivation_inefficient}

\begin{figure}%[htbp]
    \centering
    \begin{subfigure}[b]{0.235\textwidth}
        \centering
        \includegraphics[width=\textwidth]{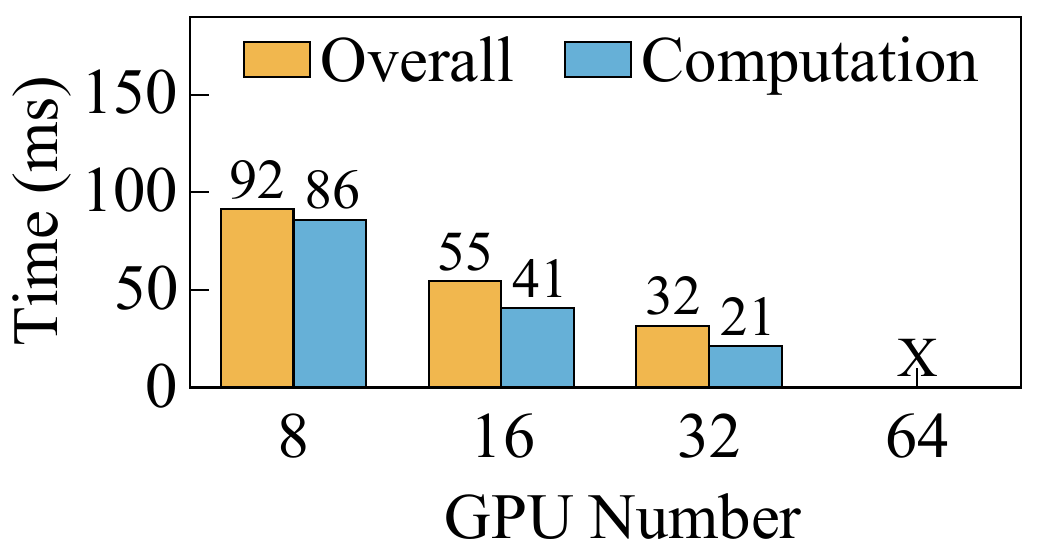}
        \subcaption{Ulyssess-Attention (MHA)}
    \end{subfigure}
    \begin{subfigure}[b]{0.235\textwidth}
        \centering
        \includegraphics[width=\textwidth]{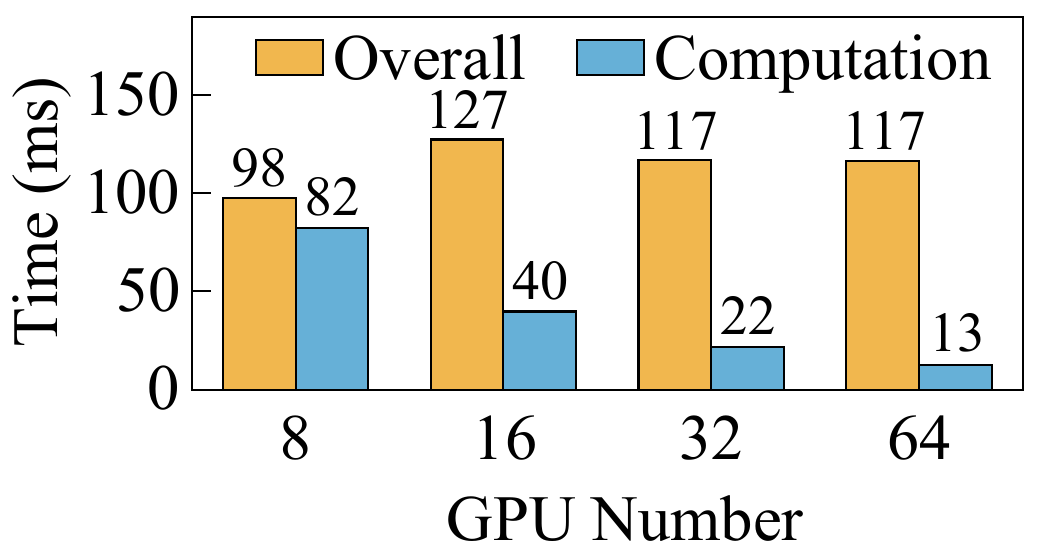}
        \subcaption{Ring-Attention (MHA)}
    \end{subfigure}
    \par\bigskip 
    \begin{subfigure}[b]{0.235\textwidth}
        \centering
        \includegraphics[width=\textwidth]{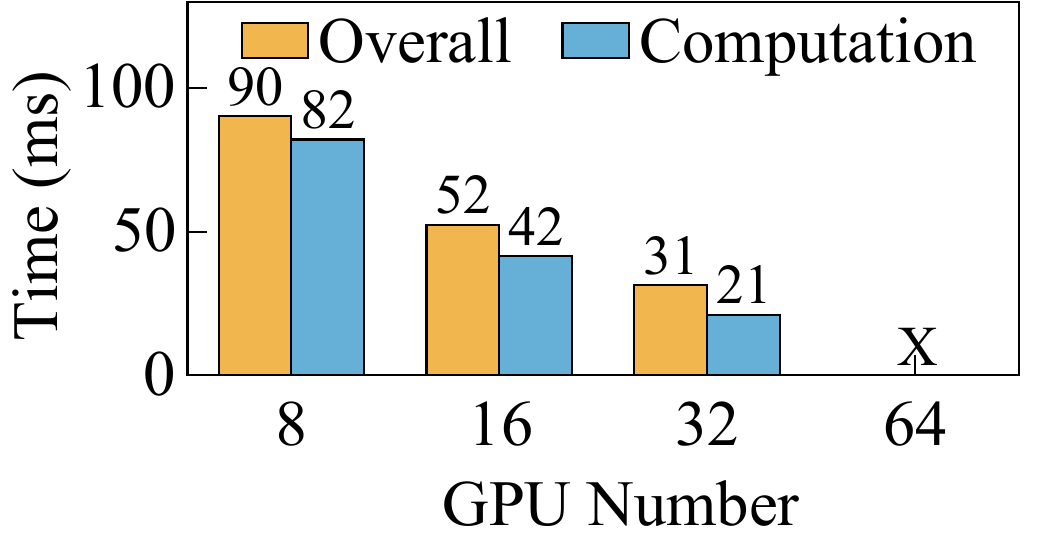}
        \subcaption{Ulyssess-Attention (GQA)}
    \end{subfigure}
        \begin{subfigure}[b]{0.235\textwidth}
        \centering
        \includegraphics[width=\textwidth]{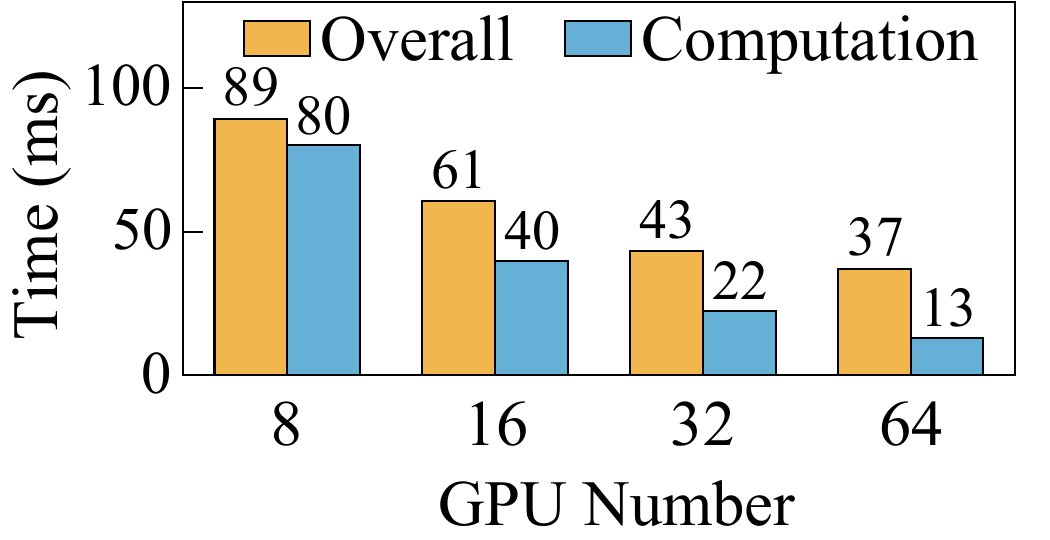}
        \subcaption{Ring-Attention (GQA)}
    \end{subfigure}
    \caption{Forward time evaluation of Ulysses-Attention and Ring-Attention on 8 physical nodes, each equipped with 8 NVIDIA Ampere GPUs connected by NVLINK. Each node has four 200 HDR NICs. In the test, we set $H=32$, $D=4096$, and $S=128K$ for MHA, and $H_{kv}=8$ for GQA.}
    \label{fig:motivation-attention-time}
\end{figure}

While Ring-Attention demonstrates the potential to scale SP to large degrees, its performance is hindered by significant communication overheads. We evaluated the performance of Ring-Attention and Ulysses-Attention with a sequence length of 128K on a testbed comprising 64 GPUs\footnote{To scale training with 1M sequence length to 2048 GPUs, constrained by the global batch size of 4M tokens, $d_{sp}$ would need to be scaled to 512. In this scenario, each query/key/value chunk contains 2K tokens, analogous to scaling the training with 128K sequence length on 64 GPUs.}. As illustrated in Figure \ref{fig:motivation-attention-time}, Ulysses-Attention and Ring-Attention exhibit similar computation time, which decreases nearly linearly with the increased number of GPUs. However, as the degree of SP increases, Ring-Attention encounters bottlenecks due to the \texttt{P2P} communication required for transferring KV chunks over the network. Specifically, with MHA, the overall execution time for Ring-Attention does not improve when scaling from 32 GPUs to 64 GPUs. Although GQA reduces the communication volume by $4\times$, Ring-Attention still takes 1.8$\times$ more time on communication than on computation when using 64 GPUs.

\begin{figure}
    \centering
    \includegraphics[width=0.465\textwidth]{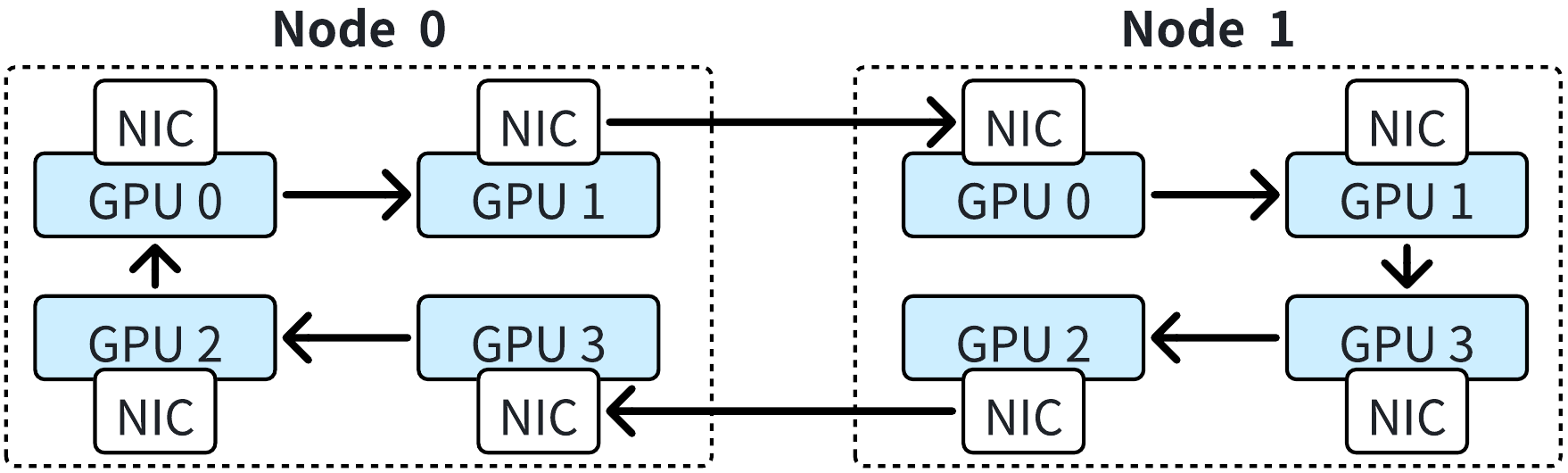}
    \caption{Ring-Attention uses one NIC for sending key/value chunks and another NIC for receiving key/value chunks.}
    \label{fig:singe_nic_problem}
\end{figure}

The performance inefficiency of Ring-Attention primarily stems from three factors. 
First, due to the small communication size, the intra-node communication via NVLINK is more sensitive to the communication latency rather than the bandwidth. When running GQA with a sequence length of 128K on 8 GPUs, the communication volume is 64MB per step. This size does not fully utilize the high bandwidth of NVLINK, resulting in high communication latency that cannot be overlapped with computation.
Second, when scaling Ring-Attention, the computation time per step decreases quadratically, whereas the communication volume per step only decreases linearly. This scaling exacerbates the imbalance between computation and communication, making communication the performance bottleneck. 
Third, Ring-Attention does not fully utilize network resources due to its ring-based communication design. Despite the widespread use of multi-rail networks in GPU clusters \cite{railonly,railarch}, Ring-Attention utilizes one NIC for sending KV chunks and another NIC for receiving KV chunks, as shown in Figure \ref{fig:singe_nic_problem}. So in a single step, all other ranks must wait for the slowest rank using inter-node \texttt{P2P} communication. Thus, it is difficult to overlap communication with computation when scaling Ring-Attention to a large scale.

%% file: 3_SystemDesign.tex
\section{Distributed 2D-Attention}
We introduce \SysName to address the scalability and efficiency challenges in training long-sequence LLMs. In particular, we propose 2D-Attention, which integrates head-parallel and context-parallel attention through a hybrid strategy, leveraging the benefits of both methods. This approach naturally overcomes the scalability limitations of head-parallel attention by incorporating context-parallel attention. To further reduce the communication overhead in Attention blocks, we design a Double-Ring-Attention mechanism and disclose the influence of device placement. Additionally, we briefly analyze the performance of 2D-Attention.

\subsection{2D-Attention Overview} \label{subsec: 2d_overview}

In \SysName, attention is parallelized across two dimensions: head parallelism (HP) and context parallelism (CP), which is referred to as \textit{2D-Attention}. It organizes $d_{sp}$ GPUs into a $d_{hp} \times d_{cp}$ grid, forming $d_{hp}$ CP process groups of size $d_{cp}$ and $d_{cp}$ HP process groups of size $d_{hp}$. Thus, we have 
$$d_{sp} = d_{hp} \times d_{cp}.$$ 
%We will discuss the selection of $d_{hp}$ and $d_{cp}$ in Section~\ref{sec:design_model}. 

\begin{figure}
    \centering
    \includegraphics[width=0.465\textwidth]{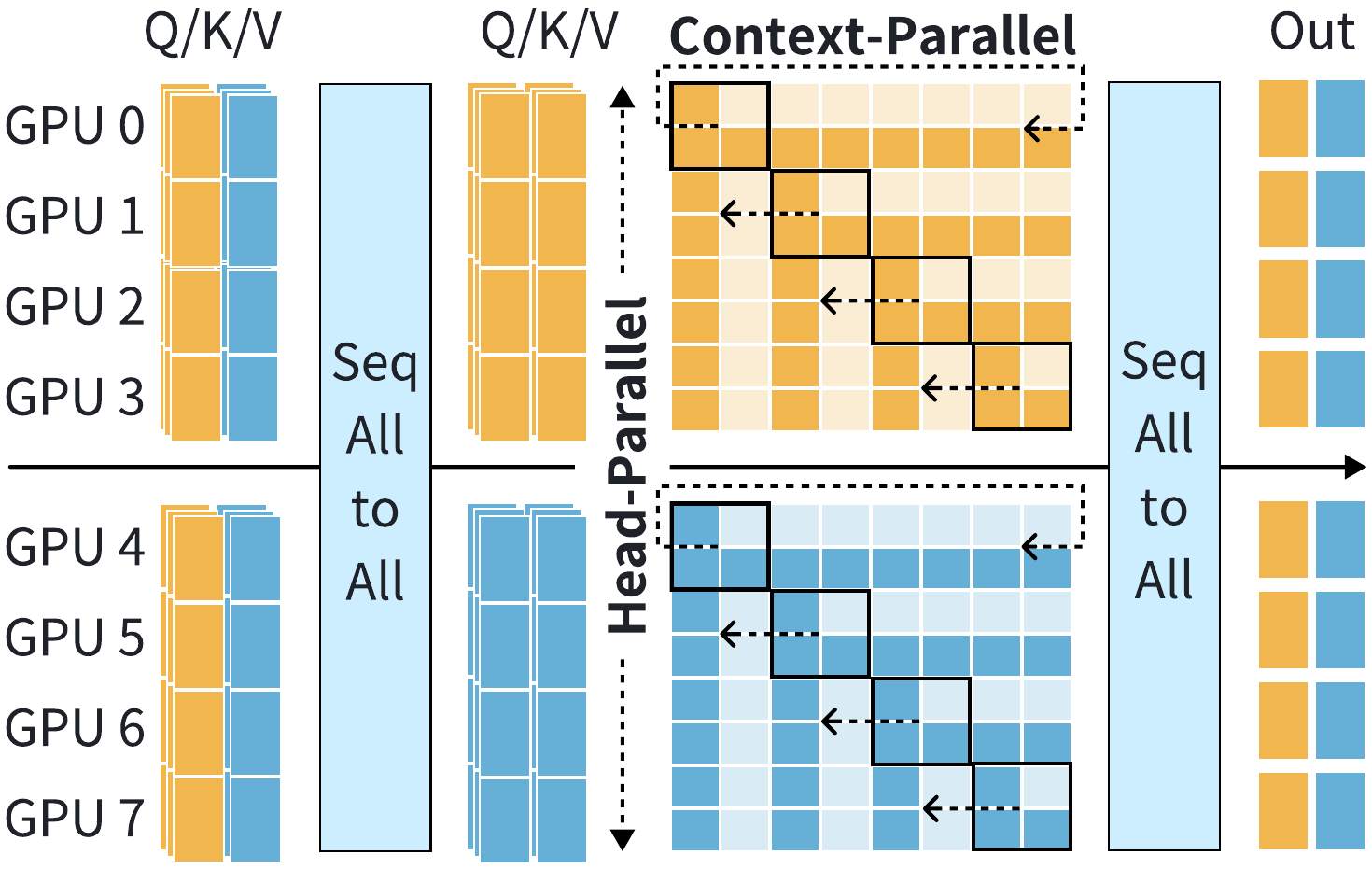}
    \caption{2D-Attention design. Different colors represent different attention heads. In this example, $d_{cp}=4$, $d_{hp}=2$.}
    \label{fig:2d_design}
\end{figure}

Algorithm~\ref{alg:2d_attention} and 
Figure~\ref{fig:2d_design}
illustrate the forward pass of 2D-Attention. In Figure~\ref{fig:2d_design}'s configuration, each CP process group contains four GPUs. The input tensors, Q (queries), K (keys), and V (values), are divided along the sequence dimension, with each segment shaped as $(H, S/d_{sp}, D/H)$.  2D-Attention handles head parallelism across CP groups, while context parallelism is executed within each CP group.

\begin{algorithm}[t]
\caption{2D-Attention Mechanism (Forward Phase)}
\label{alg:2d_attention}
\small
\begin{algorithmic}[1]
\State \textbf{Input:} $Q$, $K$, $V$, $d_{hp}$, $d_{cp}$ 
\State KV Replication: $\hat{K}, \hat{V} \gets \textbf{Replica}(K, V)$
\State Distribute QKV: $Q', K', V' \gets \textbf{SeqAlltoAll}(Q, \hat{K}, \hat{V})$
\ForAll{CP process groups}
\State $O' \gets \textbf{DoubleRingAttention}(Q', K', V', d_{cp})$
\EndFor
\State Gather output: $O \gets \textbf{SeqAlltoAll}(O')$
\State \textbf{Output:} Attention Output of shape $(H, S/d_{sp}, D/H)$
\end{algorithmic}
\end{algorithm}

The computation of MHA in 2D-Attention involves three steps. \circled{1} The \texttt{SeqAlltoAll} communication operation distributes the QKV tensors based on the head dimension across \(d_{hp}\) GPUs and re-partitions them along the sequence dimension across \(d_{cp}\) GPUs, transforming the shape of QKV to \((H/d_{hp}, S/d_{cp}, D/H)\). This step ensures that each CP group receives the entire sequence of QKV with \(H/d_{hp}\) attention heads, as illustrated in Figure~\ref{fig:2d_design}. \circled{2} Each CP group independently performs Double-Ring-Attention, as detailed in Section \ref{sec:2d_sliding_window}, resulting in an output tensor of shape \((H/d_{hp}, S/d_{cp}, \\ D/H)\). During this stage, each GPU computes attention using the local QKV and exchanges partitioned KV chunks via \texttt{P2P} communication, transferring \(2 \times (H/d_{hp}) \times (S/d_{cp}) \times (D/H) = 2SD/d_{sp}\) elements through NVLINK or network. \circled{3}  Finally, another \texttt{SeqAlltoAll}  consolidates the attention outputs across the head dimension and re-partitions the sequence dimension, transforming the output tensor to \((H, S/d_{sp}, D/H)\).

In the backward pass, a \texttt{SeqAlltoAll}  transforms the gradients of the attention output from shape $(H, S/d_{sp}, D/H)$ to $(H/d_{hp}, S/d_{cp}, D/H)$. Subsequently, each CP process group engages in context-parallel computations for the gradients by iteratively sending and receiving the partitioned KV chunks and their gradients. %The volume of data communication each GPU needs to handle is doubled compared to the forward pass. 
Finally, another \texttt{SeqAlltoAll} communication operation is employed to transform the gradients of QKV back to $(H, S/d_{sp}, D/H)$.
\subsection{KV Replication for GQA}

In MHA computation, \(d_{hp}\) can be set to up to \(H\). However, when directly computing GQA, \(d_{hp}\) is constrained by the number of KV heads \(H_{kv}\). Since \(H_{kv} < H\), this constraint limits the search space for the two-dimensional parallel strategy in 2D-Attention, potentially hindering optimal performance.

\begin{figure} 
    \centering
    \includegraphics[width=0.465\textwidth]{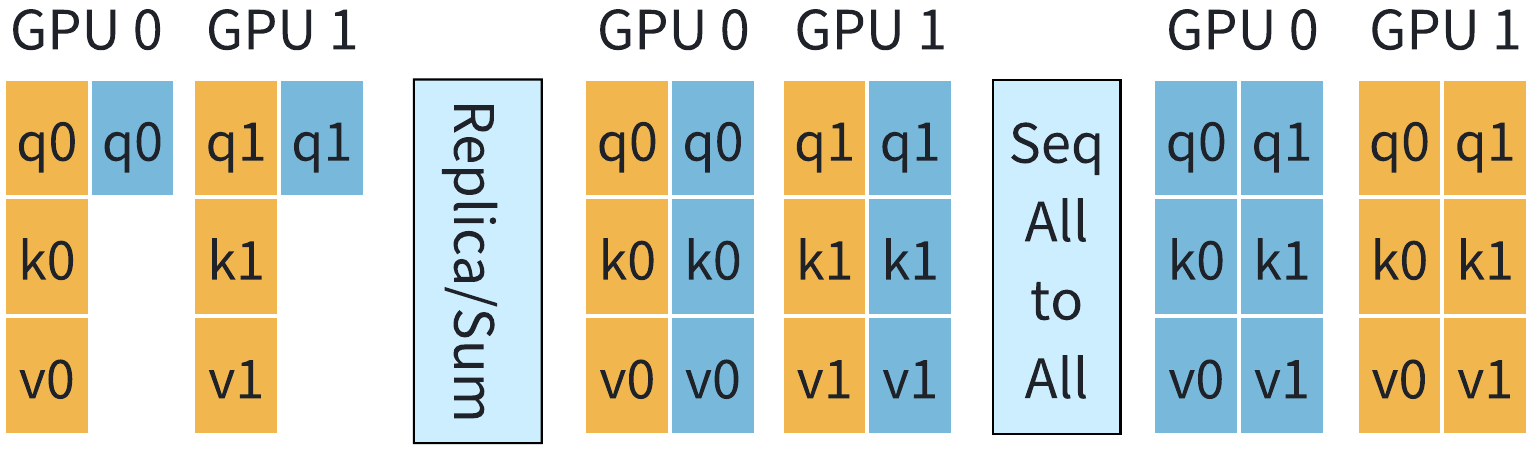}
    \caption{When $H_{kv}<d_{hp}$,  2D-Attention replicates KV tensors before \texttt{SeqAlltoAll} during forward pass, and aggregates these replicated KV tensors' gradients during backward pass.  Different colors represent different attention heads.}
    \label{fig:kv_replica_cropped}
\end{figure}

2D-Attention uses KV replication to address the constraint of limited KV heads in GQA (Figure~\ref{fig:kv_replica_cropped}). In the forward pass, the input KV tensors are shaped as \((H_{kv}, S/d_{sp}, D/H)\). To align the number of KV heads with the head-parallel size, 2D-Attention replicates KV tensors, resulting in the shape of \((\hat{H}_{kv}, S/d_{sp}, D/H)\), where \( d_{hp} \leq \hat{H}_{kv} \leq H\). A \texttt{SeqAlltoAll} operation transforms KV  to \((\hat{H}_{kv}/d_{hp}, S/d_{cp}, D/H)\).  
KV replication can potentially increase network traffic at this stage. We will analyze this impact on communication in Section~\ref{sec:design_model}.

\subsection{Double-Ring-Attention}
\label{sec:2d_sliding_window}

2D-Attention may incur high communication overhead if we directly use Ring-Attention for CP computation if the CP groups are inter-node.
As discussed in Section \ref{sec:motivation_inefficient},  Ring-Attention does not fully utilize the network resources because of its ring-based communication design. 
%This limitation hinders the ability to overlap communication with computation, posing significant challenges when scaling Ring-Attention to multiple nodes.
% For instance, when scaling $d_{sp}$ to 512 GPUs, even if \(d_{hp}\) is set to its maximum value of 32 (assuming $H=32$), each CP process group still contains 16 GPUs across two nodes (assuming 8 GPUs per node). %Inefficient communications still occur when directly using Ring-Attention.

\begin{figure}
    \centering
    \includegraphics[width=0.465\textwidth]{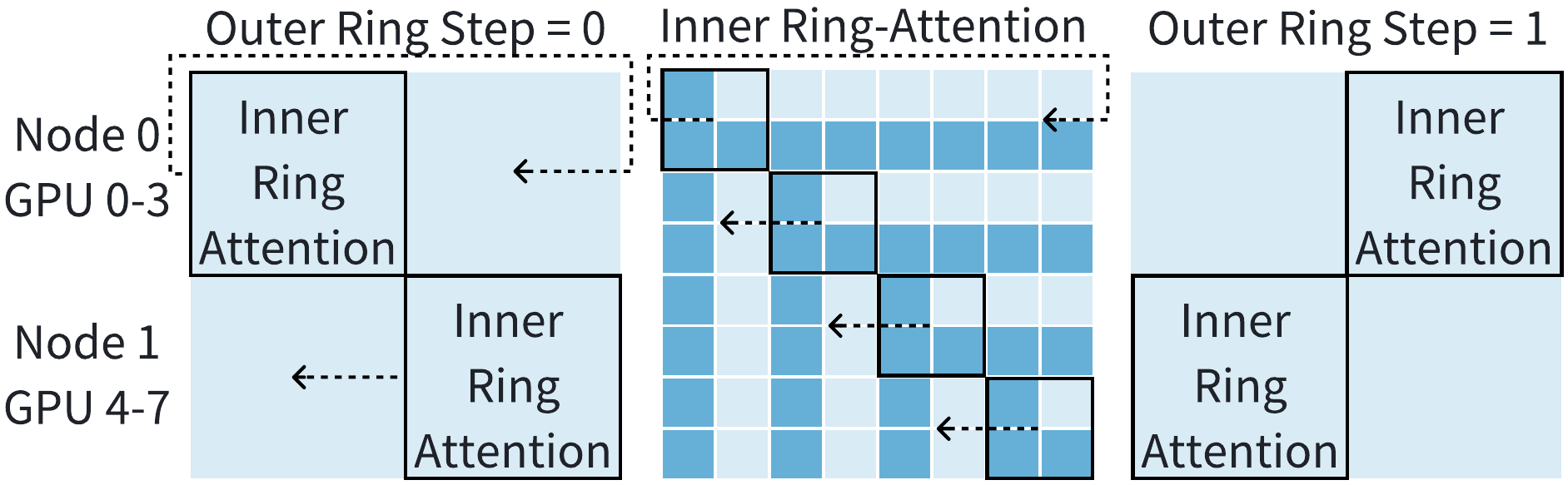}
    \caption{An illustration of Double-Ring-Attention. In this example, $d_{cp} = 8$, inner ring size is $4$ and outer ring size is $2$.}    \label{fig:sliding_window_1}
\end{figure}

To fully utilize available NICs for inter-node communication, we propose Double-Ring-Attention, which partitions the \(d_{cp}\) GPUs into multiple inner rings. As illustrated in Figure \ref{fig:sliding_window_1} and Alogrithm \ref{alg:doublering}, GPUs within each CP group form several inner rings, while the inner rings collectively form an outer ring. Assuming each inner ring consists of \(w\) GPUs, a CP process group would thus have \({d_{cp}}/{w}\) concurrent inner rings. Let \(W_{i, j}\) denote the \(j\)-th GPU in the \(i\)-th inner ring.
\circled{1} Initially, each inner ring performs conventional Ring-Attention, which involves \(w\) micro-steps. In each micro-step, a GPU performs attention computation using local QKV, while simultaneously sending and receiving KV chunks necessary for the subsequent micro-step.
\circled{2} Once the computations within all inner rings are complete, the outer ring advances to the next step and initiates a new round of Ring-Attention for each inner ring. There are \({d_{cp}}/{w}\) outer ring steps in total. In the new outer ring step, GPUs within each inner ring use new KV chunks as the initial value, fetched from GPUs of the neighboring outer ring. This \texttt{P2P} communication can be overlapped with computation: \(W_{i, j}\) sends its initial KV chunk to \(W_{i+1, j}\) and concurrently receives a KV chunk from \(W_{i-1, j}\) while computing the current inner ring.

\begin{figure}
    \centering
    \includegraphics[width=0.465\textwidth]{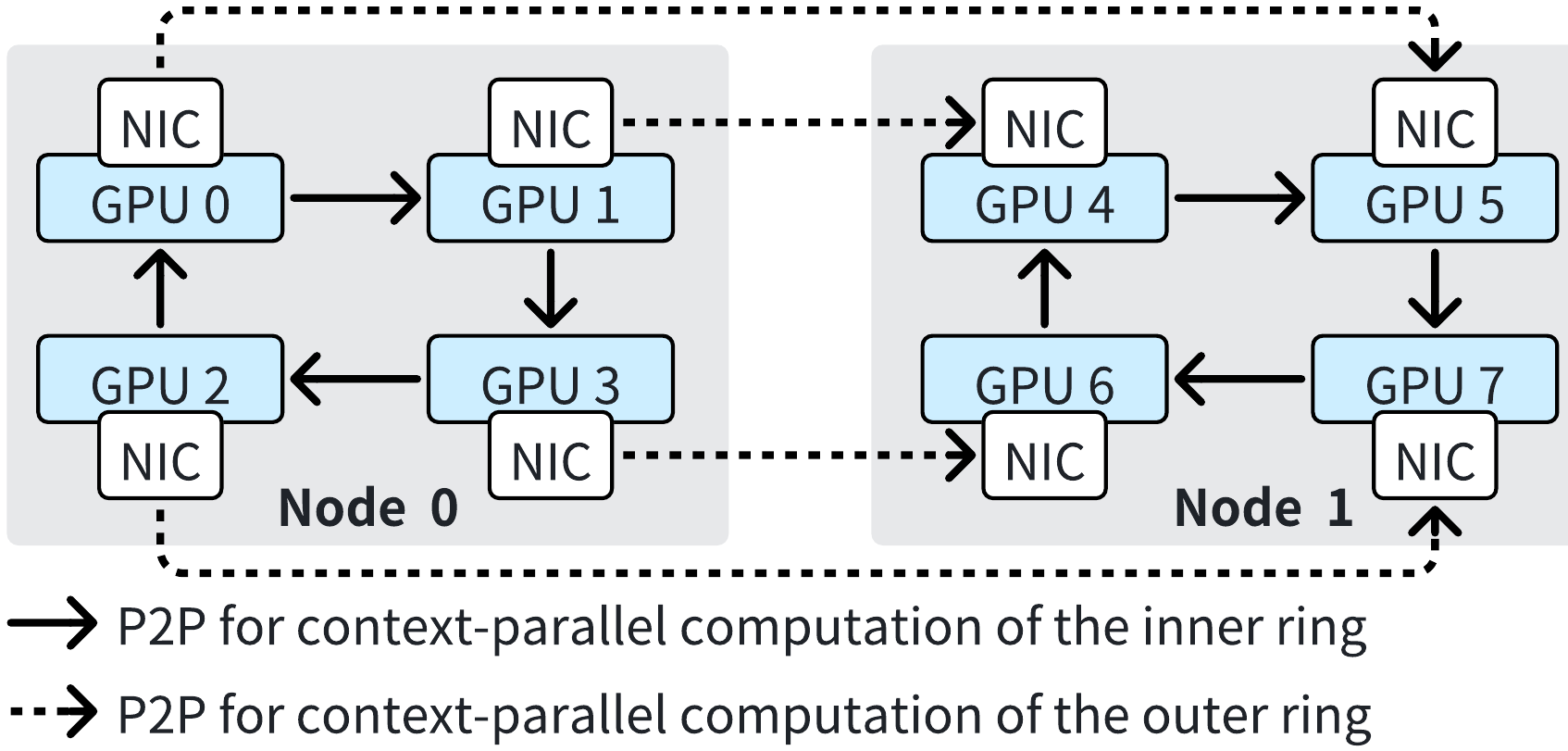}
    \caption{Communication in Double-Ring-Attention. In this example, GPUs in the same node create an inner ring with intra-node \texttt{P2P} communications. An outer ring requires inter-node \texttt{P2P} communications, utilizing all available NICs.}
    \label{fig:sliding_window_2}
\end{figure}

\begin{algorithm}
\caption{Double-Ring Attention Mechanism}
\label{alg:doublering}
\begin{algorithmic}[1]
\State \textbf{Input:} $Q$, $K$, $V$, $d_{cp}$, $w$
\For{Outer\_Ring\_Step $=0$ to $d_{cp}/w-1$} 
   \State $\text{P2P.async\_send}(KV, \text{next\_outer\_rank})$
   \State $\hat{K}, \hat{V} \gets \text{P2P.async\_recv}(\text{previous\_outer\_rank})$
   \For{Inner\_Ring\_Step $=0$ to $w-1$}
       \State $\text{P2P.async\_send}(KV, \text{next\_inner\_rank})$ 
       \State $K', V' \gets \text{P2P.async\_recv}(\text{previous\_inner\_rank})$
       \State $blocl\_out, block\_lse \gets \text{Attention($Q, K, V$)}$
       \State $out, lse \gets \text{Update($out, lse, blocl\_out, block\_lse$)}$
       \State P2P.synchronize(inner\_ring\_p2p)
       \State $K, V \gets K', V'$ \Comment{update $KV$ for next inner ring}
       \EndFor
   \State P2P.synchronize(outer\_ring\_p2p)
   \State $K, V \gets \hat{K}, \hat{V}$ \Comment{update $KV$ for next outer ring}
\EndFor
\State \textbf{Output:} $out$
\end{algorithmic}
\end{algorithm}

Double-Ring-Attention offers superior communication efficiency compared to the original Ring-Attention. It fully utilizes available network resources to transfer KV chunks across nodes and overlaps these communication processes with computational tasks. For example, in the configuration of Figure \ref{fig:sliding_window_2}, 8 GPUs are arranged into two inner rings, each containing 4 GPUs. During computation within an inner ring, GPUs 0-3 employ distinct NICs to send KV chunks to GPUs 4-7.%, which require these data in the subsequent outer ring step.
Additionally, \texttt{P2P} within the inner rings can be entirely initiated within a single node, thereby avoiding the need to wait for inter-node \texttt{P2P} communication at every micro-step. 
We will analyze the communication cost of Double-Ring-Attention and discuss the choice of $w$ in Section~\ref{sec:design_model}.

\subsection{Head-First \& Context-First Device Placement}

\begin{figure} 
    \centering
    \includegraphics[width=0.465\textwidth]{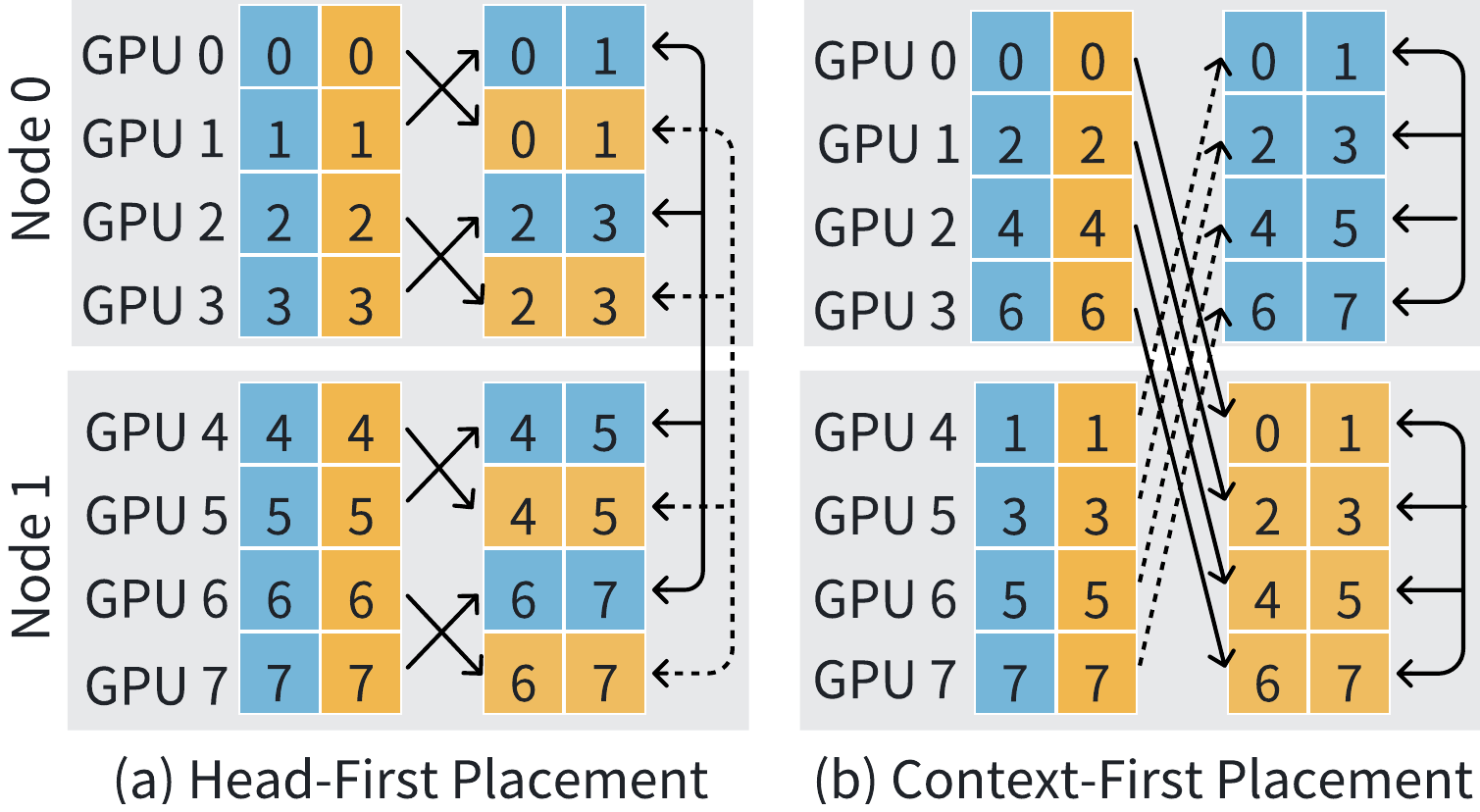}
    \caption{Context-first placement vs. head-first placement. Different colors represent different attention heads. In context-first placement, a post-processing function within the data loader is required to adjust input sequence placement at the start of each batch. }
    \label{fig:head_context_first}
\end{figure}

Given \( d_{hp} \) and \( d_{cp} \), there are two device allocation strategies: head-first placement and context-first placement. The selection of an appropriate placement strategy is critical due to the disparity between inter-node and intra-node bandwidths in GPU clusters. For instance, DGX-A100 nodes provide an intra-node bidirectional bandwidth of 600 GB/s per GPU through NVLINK, while the inter-node bidirectional bandwidth is only 400 GB/s per node.  The choice of device placement directly influences the distribution of inter-node and intra-node communication for two types of operations in 2D-Attention: \texttt{SeqAlltoAll} and \texttt{P2P}.
Figure~\ref{fig:head_context_first} shows examples of head-first and context-first placement.

In head-first placement, GPUs of the same HP group are given high priority for colocation on the same node. As illustrated in Figure~\ref{fig:head_context_first}(a), GPUs 0 and 1 are assigned to the same HP group but to different CP groups. This configuration can efficiently leverage NVLINK for \texttt{SeqAlltoAll}, as it only requires a standard NCCL  \texttt{AlltoAll} within the HP process group. %(e.g., between GPU 0 and GPU 1 in Figure~\ref{fig:head_context_first}(a)). 
However, head-first placement leads to higher inter-node traffic during Double-Ring-Attention, because GPUs within the same CP group are more likely to be distributed across different nodes, increasing the inter-node traffic.

In context-first placement, GPUs of the same CP group are prioritized for colocation on the same node. As shown in Figure~\ref{fig:head_context_first}(b), GPUs 0-3 are allocated to the same CP group. Thus, in this example, Double-Ring-Attention generates only intra-node traffic, significantly reducing the communication latency per \texttt{P2P} operation. However, when $d_{cp} > 8$, \texttt{P2P}  necessitates inter-node interconnections. Fortunately, the double-ring approach proposed in Section~\ref{sec:2d_sliding_window} leverages multiple NICs to maintain high efficiency. 
% Achieving context-first placement requires adjusting the input tensor placement across $d_{sp}$ GPUs. 
%As shown in Figure~\ref{fig:head_context_first}(b), 
Maintaining the use of a standard NCCL \texttt{AlltoAll} within an HP group necessitates reordering the input QKV tensors across nodes, which increases network traffic for each Transformer layer. To mitigate this issue, we adopt the approach used in Megatron-LM, implementing a post-processing function within the data loader to adjust input tensor placement at the start of each batch. This obviates the need for on-the-fly data movement for QKV tensors. Even with this optimization,  \texttt{SeqAlltoAll}  still demands significant inter-node communication traffic. 
%In addition, current 2D-Attention cannot overlap \texttt{SeqAlltoAll} with computation. 

% We will further analyze the influence of placement in Section~\ref{sec:design_model} and with testbed experiment results.

% The communication efficiency of context-first versus head-first placement depends on the specific values of $d_{hp}$ and $d_{cp}$.

\subsection{Performance Analysis}

\label{sec:design_model}

\subsubsection{Scalability Analysis} 
2D-Attention enhances the scalability of long-sequence training by integrating head parallelism and context parallelism through a hybrid strategy. It overcomes the limitations of head parallelism by incorporating context-parallel attention, distributing computation across a grid of GPUs organized as \(d_{hp} \times d_{cp}\). This allows sequence parallelism to scale to an unlimited number of GPUs. Additionally, in the case of GQA, 2D-Attention can scale \(d_{hp}\) to \(H\) using KV replication, ensuring flexible processing and a large search space for optimal performance.

\subsubsection{Computation Analysis.}
%\label{sec:design_perf_compt}
Given a sequence \((S, D)\), the computational complexity of attention is \(O(S^2D)\). The computation time can be formulated as \(T_{comp}^{fwd} = \alpha S^2D\), where \(\alpha\) represents the proportionality constant for the forward computation time.
In 2D-Attention, the forward computation time for each micro-step within the inner ring is described as \(\alpha \left({S}/{d_{cp}}\right)^2 {D}/{d_{hp}}.\) Since $d_{sp}=d_{hp} \times d_{cp}$, we have:
\[T_{comp}^{fwd} = \alpha {S^2D}/({d_{cp}d_{sp}}).\]
There are \(w\) micro-steps in an inner ring and \({d_{cp}}/{w}\) outer ring steps. The total forward computation time  can be expressed as: \(d_{cp} \times T_{comp}^{fwd}.\) 
For the backward pass, the computation time for each micro-step is described as:
\[T_{comp}^{bwd} = 3 \alpha {S^2D}/({d_{cp}d_{sp}}).\]
This is because the backward computation kernel naturally requires additional computations, such as activation recomputing and gradient calculations as in FlashAttention \cite{flashattn1}.

% \noindent \textbf{Insight 1:} When using 2D-Attention, HP does not introduce additional computational overhead. However, when using CP, since it launches \(d_{cp}\) computation kernels in total per layer, each kernel introduces additional time for tensor copying and transformation operations. When \(d_{cp}\) is large, this additional computational overhead cannot be neglected.
% \textcolor{red}{should be have a small trace snapshot to show this?}

\subsubsection{P2P Communication Analysis.}
%\label{sec:design_perf_p2p}
The shape of a KV chunk is defined by: $(\max({H}_{kv}, d_{hp})/d_{hp}, S/d_{cp}, D/H)$, where $H_{kv}=H$ in MHA, and  the KV tensors are replicated to match the head-parallelism size  if $d_{hp} > H_{kv}$. The size of a KV chunk can be calculated as follows: 
$$Size(kv) = \max({H}_{kv}, d_{hp}) / H \times 4SD/d_{sp},$$
where the factor of 4 accounts for two tensors with data type FP16. 
When using Double-Ring-Attention, given the inner ring size \(w\), each GPU launches \((w-1)\) \texttt{P2P} communications for the inner ring and one \texttt{P2P} communication per outer ring step (except the last one) in the forward phase. The communication size for each \texttt{P2P} communication is equivalent to $Size(kv\_chunk)$. GPUs concurrently launch \texttt{P2P} communications for inner rings and outer rings. Each \texttt{P2P} communication time depends on the slowest rank, due to the ring communication fashion.
The forward execution time per inner ring, considering the overlap between communication and computation can be formulated as follows:
\begin{equation*}
    T_{inner\_ring}^{fwd} = A \times (w - 1) + B,
\end{equation*}
where \( A \) and \( B \) are defined as:
\begin{align*}
    A = \max(T_{comp}^{fwd}, T_{P2P\_inner}^{fwd}), B = \max(T_{comp}^{fwd}, T_{P2P\_outer}^{fwd}).
\end{align*}
The backward execution time per inner ring can be expressed with similar expressions.

The per \texttt{P2P} communication time remains unaffected by \(d_{cp}\) (assuming no KV replication), as \(Size(kv\_chunk)\) remains constant regardless of \(d_{cp}\). However, the computation time per micro-step decreases linearly when \(d_{cp}\) is increased. Thus, it becomes more challenging to effectively overlap computation and communication, and  Ring-Attention exhibits poor performance in large clusters due to high communication overhead. 
2D-Attention outperforms Ring-Attention since it provides more opportunities for computation-communication overlap by limiting \(d_{cp}\).

% \textbf{Insight 1: .} 

% Since there are \(d_{cp}/w\) outer ring steps, there are a total of \((d_{cp} - 1)\) \texttt{P2P} communications. Compared to Ring-Attention, Double-Ring-Attention does not introduce additional \texttt{P2P} communications overall. Thus, the data size that each GPU sends out in the forward phase is:
% $$P2P\_Volume\_FWD = (d_{cp} - 1) \times Size(kv).$$
% During the backward phase, each GPU within a CP process group iteratively sends and receives KV chunks along with their corresponding gradients. Consequently, the data size that each GPU sends out in the backward phase is:
% $$P2P\_Volume\_BWD = 2 \times (d_{cp} - 1) \times Size(kv).$$

\noindent \textbf{Selection of Inner Ring Size.} When selecting context-first placement, ranks of the same CP group are consolidated to as few nodes as possible. In this case, there are \(w\) concurrent \texttt{P2P} communications for the outer ring. To fully utilize network resources, \(w\) should match the number of NICs per node. When \(w\) is smaller than that of NICs, we cannot fully utilize all NICs for \texttt{P2P}. Conversely, when \(w\) is larger than that of NICs, GPUs may share the same NIC for \texttt{P2P}, leading to worse performance due to congestion.

\noindent \textbf{GQA vs. MHA.} During 2D-Attention of GQA, each \texttt{P2P} transfer involves \(\hat{H}_{kv}/H \times 2SD/d_{sp}\) elements, where \(\hat{H}_{kv}\) represents the number of KV heads after KV replication. Compared to MHA, GQA requires less communication when \(\hat{H}_{kv} < H\). Specifically, when applying 2D-Attention for GQA, it results in less communication volume in the CP process group as long as \(H_{kv} < d_{hp}\), because KV replication is not applied in this case. However, if \(d_{hp} = H\), GQA and MHA will have the same communication volume due to KV replication. %Thus, MHA is more sensitive to \texttt{P2P} communication latency.
% and context-first placement is more suitable for this workload. On the contrary, head-first placement is more suitable for GQA.

% \noindent  \textbf{Insight 1:}  \SysName chooses context-first placement for MHA and head-first placement for GQA based on their communication characteristics.

% \noindent  \textbf{Insight 2:  }

\subsubsection{SeqAlltoAll Communication Analysis.}
The size of a Q chunk and output chunk can be calculated as follows:
$$Size(q) = Size(out) = {2SD}/{d_{sp}}.$$
\texttt{SeqAlltoAll} performs NCCL \texttt{AlltoAll} on \(d_{hp}\) GPUs. The size of the data that each GPU sends out in both the forward and backward phases can be expressed as follows:
$$AlltoAll\_Volume = \textstyle \sum_{i \in \{(q, k, v, out\}}Size(i) \times (d_{hp}-1)/d_{hp}.$$
With a larger \(d_{hp}\), $AlltoAll\_Volume$ increases, making the operation more substantial; if \(d_{hp} = 1\), no \texttt{SeqAlltoAll} is required but \(P2P\_Volume\) increases.
%Therefore, With higher \(d_{hp}\), the \(AlltoAll\_Volume\) increases while the \(P2P\_Volume\) decreases. 
With head-first placement, more \texttt{AlltoAll}-related traffic is carried by intra-node NVLINK, and vice versa for context-first placement.

Therefore, there is a trade-off between \(d_{cp}\) and \(d_{hp}\), as well as between the head-first and context-first placement.
\SysName's overall goal is to minimize the communication time that cannot be overlapped with computation. The problem can be formulated as:
 $$\min T_{SeqAlltoAll} + (T_{inner\_ring}^{fwd} + T_{inner\_ring}^{bwd}) \times (d_{cp}/w). $$
In the formulation, $T_{SeqAlltoAll}$ represents the \texttt{SeqAlltoAll} communication time. There are $d_{cp}/w$ inner rings to complete the execution of attention. 
% Attention computation and \texttt{P2P} can be overlapped. When they are fully overlapped, the overall time of these two parts equals to the larger one.

% In the formulation, $T_{SeqAlltoAll}$ represents the \texttt{SeqAlltoAll} communication time during the forward and backward phase. $T_{comp}$ is attention computation time and $T_{inner\_ring}$ is the \texttt{P2P} communication time. Attention computation and \texttt{P2P} can be overlapped. When they are fully overlapped, the overall time of these two parts equals to the larger one.

% $$ \max(T_{comp}^{fwd}, T_{P2P\_fwd}(inner)) \times (w-1) + T_{P2P\_fwd}(outer)$$

\subsubsection{Memory Analysis.}
When using 2D-Attention, each GPU should save its input QKV chunks (after \texttt{SeqAlltoAll}) as the activation. Thus, given a fixed sequence length, 2D-Attention can also reduce the activation memory usage by increasing \(d_{sp}\). 
%Besides, in Double-Ring-Attention, each GPU also maintains a buffer of \(Size(kv)\) for inner ring \texttt{P2P} communication and another buffer of \(Size(kv)\) for outer ring \texttt{P2P} communication. 
Similar to Ring-Attention, each GPU of \SysName maintains a buffer of \(Size(kv)\) for inner ring \texttt{P2P} communication. 
However, \SysName requires another memory buffer of \(Size(kv)\) for outer ring \texttt{P2P} communication. Experiment results in Section~\ref{sec:exp} show that this memory overhead is small and does not hinder scalability.

% \subsubsection{2D-Attention Compatibility Discussion}

\section{End-to-end System Implementation}

We describe the end-to-end system implementation of \SysName\ for training LLMs on our internal framework with two techniques: Hybrid ZeRO and \textit{selective checkpoint++}.
%We implement \SysName on our internal framework with about 2800 LOC
% In addition to 2D-Attention, we employ {ZeRO} across both sequence and data parallelism dimensions to minimize memory usage. To optimize the balance between memory consumption and computational efficiency, we introduce \textit{selective checkpoint++}, a whitelist-based mechanism for gradient checkpointing. %\todo{which framework we implement in?}

\subsection{Hybrid  ZeRO for Norm and Linear Modules}

\begin{figure}
    \centering
    \includegraphics[width=0.465\textwidth]{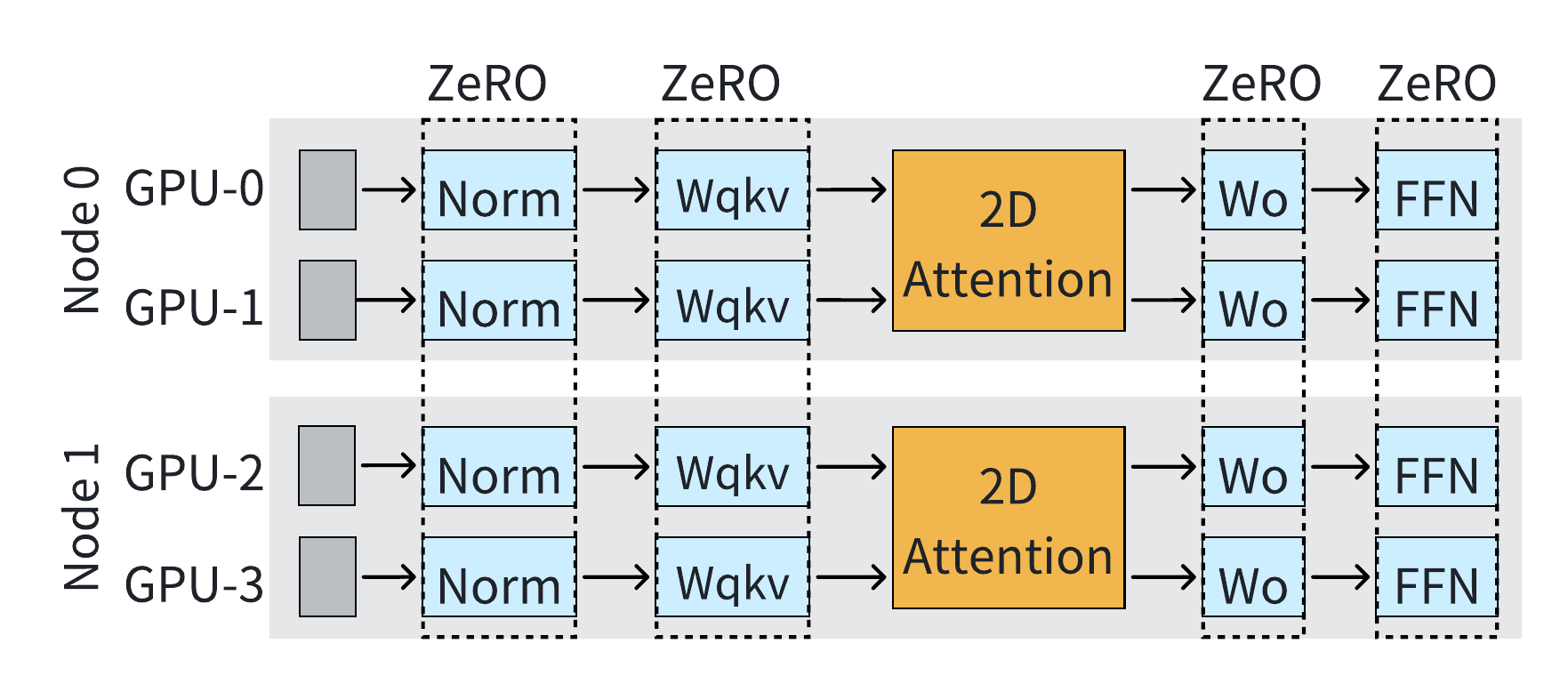}
    \caption{\SysName applies ZeRO to Norm and Linear modules across both DP and SP dimensions.}
    \label{fig:e2earch}
\end{figure}

% 2D-Attention can be integrated with existing hybrid parallelisms such as DP, TP, and PP. 
In \SysName, all modules except for attention (e.g., Linear, LayerNorm, etc.) utilize Zero~\cite{ZeRO}.  ZeRO is originally designed to reduce redundant memory usage across DP ranks. When directly using ZeRO, for instance in Figure~\ref{fig:e2earch}, it works for GPU-0 and GPU-2, as well as GPU-1 and GPU-3, which belong to the same DP group.  GPU-0 and GPU-1 would each hold half of the parameters or optimizer states, but these values remain identical, leading to redundant memory usage.

\SysName addresses these redundancies by applying ZeRO not only across the DP dimension but also along the SP dimension. This hybrid approach shards model states across both dimensions, distributing the model states across more GPUs. As a result, only ${1}/(d_{dp} \times {d_{sp}})$ of the model states are kept in GPU memory, significantly reducing the redundant memory usage. Such approach is also used in existing  frameworks like DeepSpeed-Ulysses \cite{DeepspeedUlysses}.
However, the latency of collective communication operations demonstrates a positive correlation with the communication scale \cite{sun2019gradientflow,MiCS}. Consequently, as \( d_{dp} \times d_{sp} \) scales up to hundreds of GPUs, the communication overhead becomes significant. In \SysName, we adopt the approach of AMSP \cite{amsp}, which introduces three flexible sharding strategies: Full-Replica, Full-Sharding, and Partial-Sharding. These strategies enable the Norm and Linear modules to select an appropriate sharding number across \( d_{dp} \times d_{sp} \) GPUs, effectively balancing the GPU memory usage and communication overhead.

\renewcommand{\arraystretch}{1}
\begin{table*}[]
\centering
\small
\begin{tabular}{@{}p{2cm}|*{4}{>{\raggedleft\arraybackslash}p{0.61cm}}|*{4}{>{\raggedleft\arraybackslash}p{0.61cm}}|*{4}{>{\raggedleft\arraybackslash}p{0.61cm}}|*{4}{>{\raggedleft\arraybackslash}p{0.61cm}}@{}}
\toprule
 & \multicolumn{4}{c|}{MHA (TGS)} & \multicolumn{4}{c|}{MHA (MFU)} & \multicolumn{4}{c|}{GQA (TGS)} & \multicolumn{4}{c}{GQA (MFU)} \\ \midrule
System & 128K  & 256K & 512K & 1M    & 128K  & 256K & 512K & 1M    & 128K  & 256K & 512K & 1M    & 128K  & 256K & 512K & 1M    \\ \midrule
DS-Ulysses   & 629.9 & 418.3 & 243.1 & 130.6 & 0.305 & 0.341 & 0.359 & 0.365 & 629.9 & 418.3 & 243.1 & 130.6 & 0.305 & 0.341 & 0.359 & 0.365 \\
Megatron-CP   & 296.8  & 300.0    & 260.1    & OOM    & 0.143  & 0.244  & 0.385  & OOM  & 706.2    &  476.3    & 279.6    & OOM    & 0.342  &  0.388  & 0.413  & OOM  \\ 

HP1/CP32 & 285.0 & 287.4 & 250.4 & 121.2   & 0.138 & 0.234 & 0.369 & 0.339  & 668.5 & 480.0 & 282.5 & 153.0 & 0.323 & 0.391 & 0.417 & 0.428 \\
HP2/CP16 & 311.1 & 314.9 & 267.3 & 151.6    & 0.151 & 0.256 & 0.394 & 0.423  & 740.8 & 501.3 & 290.1 & 155.9 & 0.359 & 0.408 & 0.428 & 0.436 \\
HP4/CP8  & 548.9 & 469.2 & 283.6 & \cellcolor{gray!25} 154.1   & 0.266 & 0.382 & 0.408 & \cellcolor{gray!25} 0.431   & 814.4 & 517.4 & 295.1 & 159.5 & 0.394 & 0.421 & 0.435 & 0.446 \\
HP8/CP4  & \cellcolor{gray!25} 752.4 & \cellcolor{gray!25} 498.1 & \cellcolor{gray!25} 286.1 & \cellcolor{gray!25}  154.1 & \cellcolor{gray!25} 0.364 & \cellcolor{gray!25} 0.406 & \cellcolor{gray!25} 0.418 & \cellcolor{gray!25} 0.431   & \cellcolor{gray!25} 838.1 & \cellcolor{gray!25} 528.1 & \cellcolor{gray!25} 299.5 & \cellcolor{gray!25} 160.1 & \cellcolor{gray!25} 0.406 & \cellcolor{gray!25} 0.430 & \cellcolor{gray!25} 0.442 & \cellcolor{gray!25} 0.448 \\
HP16/CP2 & 714.3 & 472.4 & 278.9 & 150.9 & 0.346 & 0.385 & 0.412 & 0.422 & 771.4 & 498.6 & 288.0 & 155.1   & 0.373 & 0.406 & 0.425 & 0.433   \\
HP32/CP1 & 700.1 & 459.3 & 268.8 & 146.0  & 0.339 & 0.374 & 0.397 & 0.408  & 717.1 & 468.4 & 262.4 & 147.5   & 0.347 & 0.381 & 0.387 & 0.412
\\ 
\bottomrule
\end{tabular}
\caption{Performance comparison of end-to-end training between \SysName, DS-Ulysses, and Megatron-CP. HP$n$/CP$m$ denotes our proposed system \SysName (head-first placement) with head parallelism size $n$ and context parallelism size $m$. 
% The highest TGS and MFU values for each sequence length are highlighted.
}
\label{tab:compare_with_sota_sys}
\end{table*}
\renewcommand{\arraystretch}{1}

\subsection{Selective Checkpoint++}

Long sequence training leads to significant memory costs, making gradient checkpointing a common practice. During forward propagation, the gradient checkpointing mechanism stores only the input tensors of the wrapped function by the \texttt{checkpoint} function. If the dropped activation values are needed during backward propagation, they are recomputed. Typically, when we wrap the \texttt{checkpoint} function around an entire Transformer layer, the total memory required for activations of a Transformer layer is \(2SD/d_{sp}\) in FP16.

While saving the checkpoints of the entire model significantly reduces the memory footprint, it introduces additional computation overhead~\cite{flashattn1}. Given that the recomputation time for attention blocks is particularly long, a straightforward approach is to keep the activations of attention blocks and use checkpointing for the other parts of the model selectively with the provided APIs \cite{Nvidia3}. However, this solution is not memory-efficient. During backward propagation, each attention block requires extra memory to save the QKV tensors (size \(6SD/d_{sp}\) in FP16) and \texttt{softmax\_lse} (size \(4SH/d_{sp}\) in FP32) \cite{chen2024internevo}. 
To reduce memory usage, DistFlashAttn~\cite{lightseq} places the attention module at the end of each Transformer layer. This strategy eliminates the need to recompute the attention module during the backward phase and only requires storing the output of the attention module.

% \begin{figure}
%     \centering
%     \includegraphics[width=0.465\textwidth]{figures/selective_cropped.pdf}
%     \caption{Comparison between directly applying \textit{checkpoint} API and selective checkpoint++.}
%     \label{fig:selective}
% \end{figure}

\SysName implements the \textit{selective checkpoint++} mechanism without modifying the model structure. It adds attention modules to a \textit{whitelist}. During the forward pass, when encountering a module in the \textit{whitelist}, the modified \texttt{checkpoint} function saves its outputs. Specifically, for attention, it saves the attention output with the size of $2SD/d_{sp}$ and \texttt{softmax\_lse} with the size of $4SH/d_{sp}$. During the backward pass, when encountering a module in the \textit{whitelist}, the \texttt{checkpoint} function does not perform recomputation. Instead, it retrieves the stored outputs and continues the computation graph. This eliminates the need to recompute attention during the backward pass, requiring an additional $(2SD+4SH)/d_{sp}$ memory size per Transformer layer. Additionally, \textit{selective checkpoint++} is compatible with other offload techniques \cite{ren2021zero}, which involve offloading attention outputs to memory or NVMe storage.

% \noindent\textbf{Comparison with DistFlashAttn.} DistFlashAttn~\cite{lightseq} places the attention module at the last within each Transformer layer, eliminating the need for recomputing the attention module during the backward phase. However, DistFlashAttn requires modifying the model architecture, which may not be practical in many scenarios. In contrast, the selective checkpoint++ mechanism does not require modifications to the model architecture and can be easily adapted to other model architectures by adding specified modules to the \textit{whitelist}.

% Saving checkpoints for the whole LLM model significantly reduces memory footprint but brings more computation overhead~\cite{flashattn1}. As the recomputation time for attention blocks (MHA or GQA) is too long, a straightforward approach is to keep the activations of attention blocks and save checkpoints for the other model parts with the APIs provided by deep learning frameworks (Figure~\ref{fig:selective}(a)). 
% However, this is still not memory-efficient enough. For each attention block, the backward propagation computation requires extra memory size of at least $3BSD$ for saving the QKV tensors and \texttt{softmax\_lse}. 

%% file: 4_Evaluation.tex
\section{Performance Evaluation}
\label{sec:exp}

\subsection{Experiment Setup}

\textbf{Testbed.} We conduct performance evaluation on a cluster with 8 GPU servers unless specified otherwise. Each server is equipped with 8 NVIDIA Ampere GPUs, 128 CPU cores, and 80GB memory per GPU. Within each node, GPUs are interconnected via NVLINK. Inter-node communication is facilitated by 4 NVIDIA Mellanox HDR (200Gb/s) InfiniBand NICs, without SHARP.

\noindent \textbf{System Configurations. } We evaluate the training performance of \SysName using the configuration of LLaMA2-7B \cite{LLaMA2}, where \( D=4096 \) and \( H=32 \) for MHA, and \( H_{kv}=8 \) for GQA. The input sequence length is scaled from 128K to 1M. In all experiments, activation checkpointing is enabled by default. We analyze the performance of \SysName with different parallelism settings and device placements.

\noindent \textbf{Evaluation Metrics.} We focus on key performance metrics, including Model FLOPs Utilization (MFU) \cite{palm} and Tokens per GPU per Second (TGS). We use the formula provided in Megatron-LM \cite{Megatron-LM} for calculating FLOPs and MFU. Notably, the FLOPs for attention are halved in this work to account for the causal mask, which reduces the number of elements in attention that require computation by approximately half. This differs from the FLOPs and MFU calculations used in other works \cite{chen2024internevo, flashattn1, dao2023flashattention}, but is essential since attention accounts for the majority of the workload in long sequence training. Without this adjustment, the MFU can exceed 1, misrepresenting the actual system performance.

\noindent \textbf{Baselines.} We compare the performance of \SysName  against two long sequence training frameworks: DeepSpeed-Ulysses (DS-Ulysses) \cite{DeepspeedUlysses} and Megatron Context Parallelism (Megatron-CP) \cite{megatroncp}. DS-Ulysses employs head-parallel attention, while Megatron-CP utilizes Ring-Attention with load balancing. All baseline systems are integrated with FlashAttention-V2 \cite{dao2023flashattention}. The versions used are as follows: 1) DS-Ulysses: DeepSpeed V0.14.0; 2)  Megatron-CP: Nemo v2.0.0rc0, NemoLauncher v24.05, Megatron-Core v0.7.0, TransformerEngine v1.6, Apex commit ID 810ffa.

\renewcommand{\arraystretch}{1}
\begin{table*}[]
\centering
\small
\begin{tabular}{@{}p{0.05cm}p{0.3cm}p{0.3cm}|*{4}{>{\raggedleft\arraybackslash}p{0.65cm}}|*{4}{>{\raggedleft\arraybackslash}p{0.65cm}}|*{4}{>{\raggedleft\arraybackslash}p{0.65cm}}|*{4}{>{\raggedleft\arraybackslash}p{0.65cm}}@{}}
\toprule
& &    & \multicolumn{4}{c|}{128K}   & \multicolumn{4}{c|}{256K}  & \multicolumn{4}{c|}{512K}   & \multicolumn{4}{c}{1M}   \\
% \cmidrule(lr){4-7} \cmidrule(lr){8-11} \cmidrule(lr){12-15} \cmidrule(lr){16-19}
\midrule
& &    & \multicolumn{2}{c}{With SC++} & \multicolumn{2}{c|}{W/O SC++} & \multicolumn{2}{c}{With SC++} & \multicolumn{2}{c|}{W/O SC++} & \multicolumn{2}{c}{With SC++} & \multicolumn{2}{c|}{W/O SC++} & \multicolumn{2}{c}{With SC++} & \multicolumn{2}{c}{W/O SC++} \\
% \cmidrule(lr){4-7} \cmidrule(lr){8-11} \cmidrule(lr){12-15} \cmidrule(lr){16-19}
\midrule
& $d_{cp}$ & $d_{hp}$ & HF  & CF  & HF & CF & HF   & CF   & HF  & CF  & HF   & CF   & HF & CF  & HF   & CF  & HF & CF           \\
\midrule
\multirow{6}{*}{\rotatebox{90}{MHA}} 
& 64 & 1  & 0.092 & 0.092 & 0.070 & 0.070 & 0.159 & 0.159 & 0.122 & 0.122 & 0.290 & 0.290 & 0.221 & 0.221 & 0.452 & 0.452 & 0.357 & 0.357 \\
& 32 & 2  & 0.099 & 0.158 & 0.077 & 0.126 & 0.173 & 0.278 & 0.133 & 0.219 & 0.316 & 0.434 & 0.243 & 0.353 & 0.475 & 0.486 & 0.394 & 0.406 \\
& 16 & 4  & 0.176 & 0.245 & 0.141 & 0.205 & 0.314 & 0.378 & 0.248 & 0.317 & 0.470 & 0.472 & 0.384 & 0.388 & 0.520 & 0.509 & 0.418 & 0.413 \\
& 8  & 8  & 0.283 & 0.321 & 0.236 & 0.282 & 0.434 & 0.420 & 0.361 & 0.357 & \cellcolor{gray!25} 0.502 & \cellcolor{gray!25} 0.478 & \cellcolor{gray!25} 0.409 & \cellcolor{gray!25} 0.394 & \cellcolor{gray!25} 0.527 & \cellcolor{gray!25} 0.521 & \cellcolor{gray!25} 0.424 & \cellcolor{gray!25} 0.420 \\
& 4  & 16 & \cellcolor{gray!25} 0.328 & 0.327 & \cellcolor{gray!25}  0.289 & 0.283 & \cellcolor{gray!25} 0.436 & \cellcolor{gray!25} 0.423 & \cellcolor{gray!25} 0.369 & \cellcolor{gray!25} 0.359 & 0.487 & 0.476 & 0.399 & \cellcolor{gray!25} 0.394 & 0.519 & 0.520 & 0.418 & 0.412 \\
& 2  & 32 & 0.320 & \cellcolor{gray!25} 0.329 & 0.284 & \cellcolor{gray!25} 0.293 & 0.421 & 0.421 & 0.353 & 0.357 & 0.474 & 0.478 & 0.388 & 0.394 & 0.517 & 0.517 & 0.415 & 0.406 \\

\hline
\multirow{6}{*}{\rotatebox{90}{GQA}} & 64 & 1  & 0.255 & 0.255 & 0.196 & 0.196 & 0.379 & 0.379 & 0.308 & 0.308 & 0.470 & 0.470 & 0.378 & 0.378 & 0.508 & 0.508 & 0.406 & 0.406 \\
& 32 & 2  & 0.283 & 0.317 & 0.233 & 0.269 & 0.419 & 0.429 & 0.345 & 0.354 & 0.492 & 0.485 & 0.398 & 0.392 & 0.521 & 0.516 & 0.418 & 0.416 \\
& 16 & 4  & 0.354 & 0.338 & 0.309 & 0.294 & 0.466 & 0.437 & 0.385 & 0.373 & 0.505 & 0.494 & 0.410 & 0.404 & 0.531 & 0.526 & 0.425 & 0.426 \\
& 8  & 8  & \cellcolor{gray!25} 0.377 & \cellcolor{gray!25} 0.354 & \cellcolor{gray!25} 0.327 & \cellcolor{gray!25}  0.310 & \cellcolor{gray!25} 0.480 & \cellcolor{gray!25} 0.452 & \cellcolor{gray!25} 0.392 & \cellcolor{gray!25} 0.380 & \cellcolor{gray!25} 0.516 & \cellcolor{gray!25} 0.502 & \cellcolor{gray!25} 0.419 & \cellcolor{gray!25} 0.412 & \cellcolor{gray!25} 0.543 & \cellcolor{gray!25} 0.536 & \cellcolor{gray!25} 0.435 & \cellcolor{gray!25} 0.432 \\
& 4  & 16 & 0.354 & 0.341 & 0.310 & 0.308 & 0.457 & 0.437 & 0.377 & 0.373 & 0.500 & 0.493 & 0.409 & 0.405 & 0.532 & 0.529 & 0.428 & 0.419 \\
& 2  & 32 & 0.323 & 0.333 & 0.285 & 0.295 & 0.424 & 0.422 & 0.349 & 0.360 & 0.476 & 0.481 & 0.389 & 0.394 & 0.518 & 0.518 & 0.415 & 0.406 \\
\bottomrule
\end{tabular}
\caption{End-to-end training performance (MFU) of 7B-MHA and 7B-GQA on 64 GPUs with $d_{sp}=64$. SC++ stands for Selective Checkpoint++, HF for head-first, and CF for context-first. The highest MFU value in each column is highlighted.}
\label{tab:e2e_mha_gqa_mfu}
\end{table*}
\renewcommand{\arraystretch}{1}

\begin{figure}%[htbp]
    \centering
    \begin{subfigure}[b]{0.48\textwidth}
        \centering
        \includegraphics[width=\textwidth]{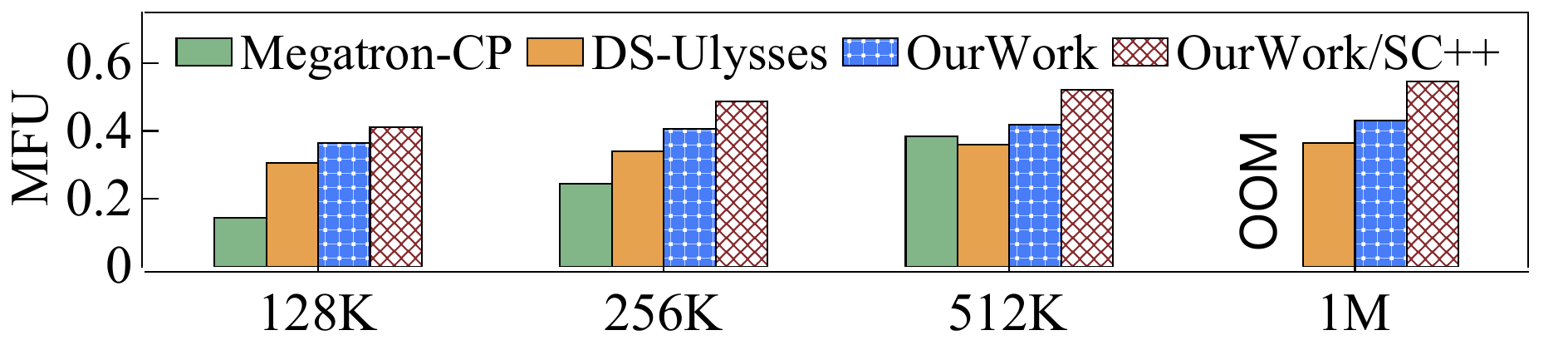}
        \caption{MHA}
    \end{subfigure}
            \par\bigskip 
    \begin{subfigure}[b]{0.48\textwidth}
        \centering
        \includegraphics[width=\textwidth]{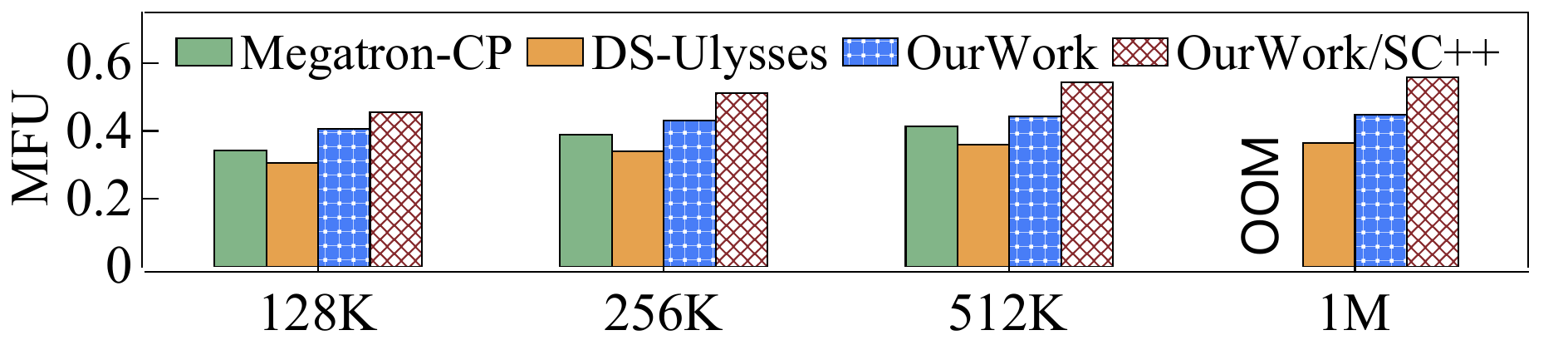}
        \caption{GQA}
    \end{subfigure}
    \caption{Performance comparison between Megatron-CP, DeepSpeed-Ulysses and our proposed \SysName on 32 GPUs with the sequence length from  128K to 1M.}
    \label{fig:e2e_compare}
\end{figure}

\subsection{Comparison with DS-Ulysses \& Megatron-CP} 
Theoretically, 2D-Attention when \( d_{cp} = 1 \) is equivalent to DS-Ulysses and 2D-Attention when \( d_{hp} = 1 \) is equivalent to Megatron-CP. 
To validate that our \SysName implementation is consistent with this theoretical analysis, we measured the TGS and MFU when training 7B-MHA and 7B-GQA on 32 GPUs using \SysName, DS-Ulysses, and Megatron-CP, with different sequence lengths. The comparison was limited to 32 GPUs because DS-Ulysses supports only head-parallelism, which is constrained by the number of attention heads. To ensure a fair comparison, all systems applied ZeRO-1 on Norm and Linear modules across the 32 GPUs, and did not use \textit{Selective Checkpoint++}. The results are shown in Table~\ref{tab:compare_with_sota_sys}.

When \( d_{cp} = 1 \), \SysName outperforms DS-Ulysses due to its superior overlap capability between communication and computation during the backward phase for Norm and Linear modules. When \( d_{hp} = 1 \), \SysName demonstrates slightly lower performance than Megatron-CP in MHA, but exhibits higher performance in GQA. Our analysis indicates both systems perform similarly in attention computation. The main performance disparity arises from the divergent choices in computation and communication operators. Notably, when processing the sequence length of 1M, Megatron-CP encounters out-of-memory errors due to increased pre-allocated GPU memory requirements for parameters and gradients.

For sequence lengths of 128K and 256K, Megatron-CP  exhibits poor performance in MHA, as the \texttt{P2P} communication cannot be effectively overlapped with computation. However, with the sequence lengths of 512K and 1M, both Megatron-CP and \SysName-HP1/CP32 show better performance than DS-Ulysses for MHA.  Additionally, in GQA, the communication volume per micro-step is reduced by a factor of 4. Consequently, Megatron-CP and \SysName-HP1/CP32 consistently outperform DS-Ulysses across all evaluated sequence lengths for GQA.

Then, we compare the end-to-end performance of the complete \SysName and the baselines.
All of the techniques such as hybrid ZeRO and Selective Checkpoint++ are used.
As shown in Figure~\ref{fig:e2e_compare},
\SysName delivers larger MFU.
The configuration of \(d_{hp}=8\) and \(d_{cp}=4\) is  more efficient in this experiment. Compared to DS-Ulysses, \SysName improves the training performance of MHA and GQA by up to $1.49\times$ and $1.53\times$, respectively. Compared to Megatron-CP, \SysName enhances the performance of MHA and GQA by up to $2.88\times$ and $1.33\times$, respectively.

%\subsection{E2E Performance with Selective Checkpoint++}
\subsection{Analysis of \SysName Performance}
\label{sec:eval_e2e_e2e}
\begin{figure}%[htbp]
    \centering
    \begin{subfigure}[b]{0.48\textwidth}
        \centering
        \includegraphics[width=\textwidth]{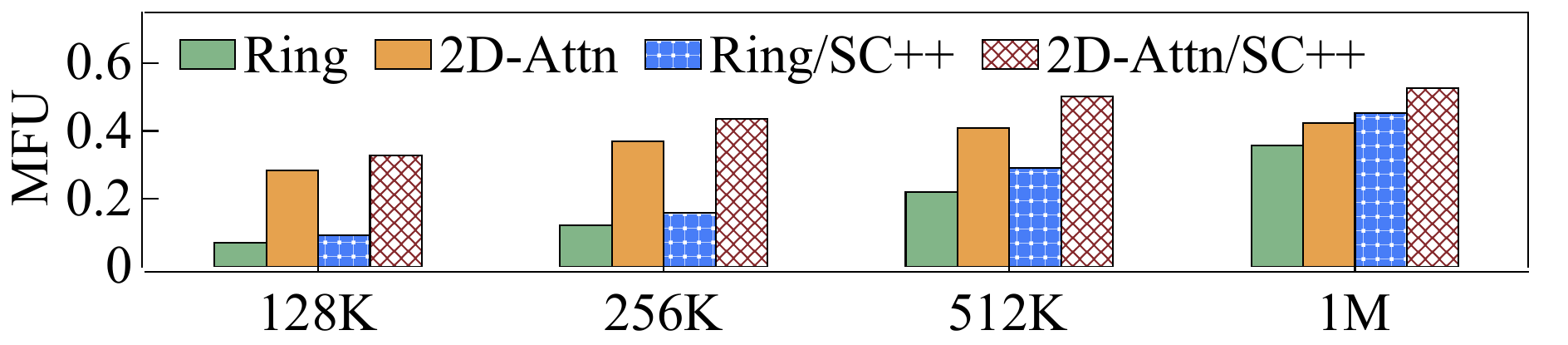}
        \caption{MHA}
    \end{subfigure}
            \par\bigskip 
    \begin{subfigure}[b]{0.48\textwidth}
        \centering
        \includegraphics[width=\textwidth]{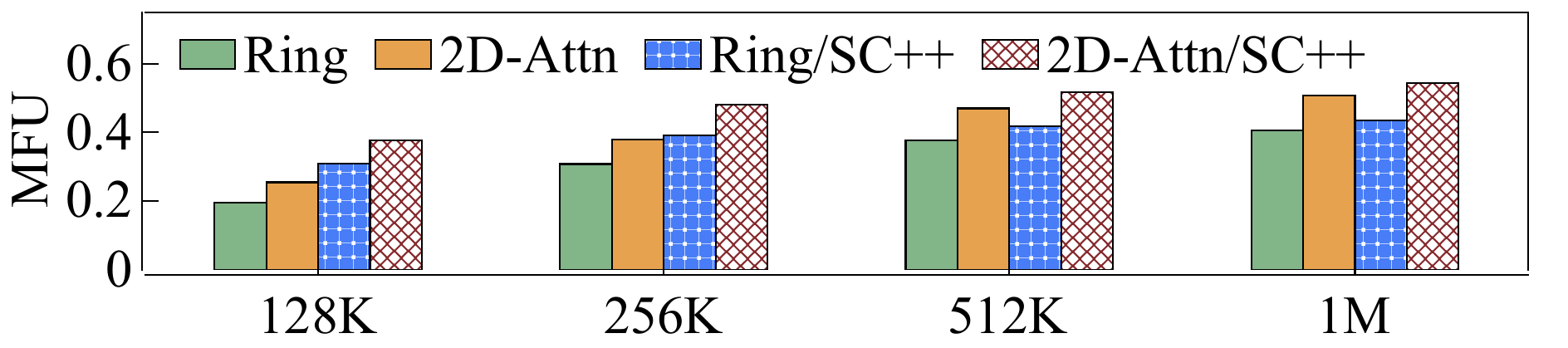}
        \caption{GQA}
    \end{subfigure}
    \caption{MFU comparison on 64 GPUs with sequence lengths from 128K to 1M. Ring indicates $d_{hp}=1$ in \SysName. 2D-Attn indicates the best-performing configuration.}    
    \label{fig:e2e_compare_64GPU}
\end{figure}

To analyze how much performance improvement can be brought by each design, 
we evaluated the performance of \SysName for training the 7B-MHA and 7B-GQA models on 64 GPUs with various sequence lengths and configurations. The evaluation results are presented in Table~\ref{tab:e2e_mha_gqa_mfu}. We do not show the results for \( d_{cp} = 1 \) as \( d_{hp} \) cannot exceed the number of attention heads, which is 32. The end-to-end evaluation demonstrates that \SysName's designs (e.g., 2D-Attention) and implementation techniques (e.g., Selective Checkpoint++), significantly enhance the training performance across all cases. Figure~\ref{fig:e2e_compare_64GPU} shows the end-to-end MFU results and the details are listed in Table~\ref{tab:e2e_mha_gqa_mfu}.

\renewcommand{\arraystretch}{1}
\begin{table*}[htbp]
\centering
\small
\begin{tabular}{@{}p{0.05cm}p{0.3cm}p{0.3cm}|*{4}{>{\raggedleft\arraybackslash}p{0.65cm}}|*{4}{>{\raggedleft\arraybackslash}p{0.65cm}}|*{4}{>{\raggedleft\arraybackslash}p{0.65cm}}|*{4}{>{\raggedleft\arraybackslash}p{0.65cm}}@{}}
\toprule
 & &  & \multicolumn{4}{c|}{MHA (Head-First)} & \multicolumn{4}{c|}{MHA (Context-First)} & \multicolumn{4}{c|}{GQA (Head-First)} & \multicolumn{4}{c}{GQA (Context-First)} \\ \midrule
& $d_{cp}$ & $d_{hp}$ & 128K  & 256K  & 512K  & 1M & 128K  & 256K  & 512K  & 1M & 128K  & 256K  & 512K  & 1M & 128K  & 256K  & 512K  & 1M \\ \midrule
\multirow{6}{*}{\rotatebox{90}{Overall}} & 64 & 1  & 296.4 & 597.8 & 1210 & 2897 & 296.4 & 597.8 & 1210 & 2897 & 86.0  & 225.1 & 713.5 & 2681 & 86.0  & 225.1 & 713.5 & 2681 \\
& 32 & 2  & 273.6 & 546.8 & 1106 & 2745 & 162.4 & 328.7 & 782.5  & 2663 & 75.4  & 198.5 & 679.5 & 2607 & 64.9  & 187.1 & 683.5 & 2589 \\
& 16 & 4  & 137.0 & 275.8 & 708.1  & 2595 & 87.4  & 213.8 & 691.5  & 2617 & 55.4  & 172.1 & 659.4 & 2559 & 59.9  & 179.1 & 668.3 & 2543 \\
& 8  & 8  & 72.2  & 187.9 & \cellcolor{gray!25} {658.3}  & \cellcolor{gray!25} 2557 & 62.2  & 185.6 & \cellcolor{gray!25} 675.3  & \cellcolor{gray!25} {2539}  & \cellcolor{gray!25} {52.1}  & \cellcolor{gray!25} {166.2} & \cellcolor{gray!25} {644.1} & \cellcolor{gray!25} {2494} & \cellcolor{gray!25} 56.8   & \cellcolor{gray!25} 175.2 & \cellcolor{gray!25} 656.1 & \cellcolor{gray!25} 2495 \\
& 4  & 16 & \cellcolor{gray!25} {58.4}  & \cellcolor{gray!25} {179.8} & 671.9  & 2575 &  60.1  & \cellcolor{gray!25} 182.6 &  680.6    & 2549 & 55.8  & 173.6 & 659.6 & 2530 & 57.3  & 177.2 & 661.7 & 2510 \\
& 2  & 32 & 60.8 & 186.0 & 684.9  & 2573 & \cellcolor{gray!25} 59.4  & 183.0 & 677.1  & 2553 & 60.8  & 185.8 & 683.9 & 2579 & 59.3  & 183.1 & 677.5 & 2555 \\
\hline
\multirow{6}{*}{\rotatebox{90}{SeqAlltoAll}} & 64 & 1  & 0.00  & 0.00  & 0.00   & 0.00   & 0.00  & 0.00  & 0.00   & 0.00   & 0.00  & 0.00  & 0.00  & 0.00   & 0.00  & 0.00  & 0.00  & 0.00   \\
& 32 & 2  & 2.23  & 3.20  & 5.49   & 10.00  & 7.19  & 13.27 & 25.10  & 49.26  & 1.89  & 2.51  & 3.92  & 6.58   & 4.92  & 8.65  & 16.29 & 31.59  \\
& 16 & 4  & 2.45  & 3.52  & 5.80   & 10.53   & 10.31 & 19.25 & 37.37  & 73.74  & 2.15  & 2.76  & 4.08  & 6.76   & 6.90  & 12.55 & 23.82 & 46.87  \\
& 8  & 8  & 3.00  & 4.15  & 6.27   & 11.22  & 12.05 & 22.26 & 42.82  & 83.30  & 2.64  & 3.24  & 4.43  & 7.31   & 8.13  & 14.60 & 27.51 & 53.35  \\
& 4  & 16 & 9.11  & 15.99 & 29.02  & 55.38  & 12.95 & 23.97 & 45.52  & 90.28  & 7.23  & 12.85 & 22.56 & 42.44  & 10.12 & 18.91 & 34.94 & 71.51  \\
& 2  & 32 & 13.42 & 23.43 & 42.73  & 81.47  & 14.56 & 25.41 & 48.25  & 100.0 & 13.40 & 23.35 & 42.85 & 81.76  & 14.31 & 25.75 & 48.43 & 106.8 
\\ \bottomrule
\end{tabular}
\caption{Average overall execution time (ms) and SeqAlltoAll time (ms) of a single 2D-Attention forward and backward operation on 64 GPUs with $d_{sp}=64$. The lowest overall execution time in each column is highlighted.}
\label{tab:overall_a2a_time}
\end{table*}
\renewcommand{\arraystretch}{1}

When \( d_{hp} = 1 \), \SysName exhibits similarly poor performance as Ring-Attention for MHA: the MFU is less than 10\% with the sequence length of 128K. When the sequence length increases to 1M, which entails a higher computational workload, the MFU is only 35.7\% without Selective Checkpoint++. For GQA, Ring-Attention involves $4\times$ less communication volume compared to MHA, leading to a higher MFU than MHA. Specifically in Ring-Attention, the MFU reaches 19.6\% with the sequence length of 128K, and increases to 40.6\% when the sequence length is 1M.

With 2D-Attention, \SysName significantly improves the training performance for MHA. Compared to Ring-Attention, 2D-Attention enhances the MFU by 4.1$\times$, 3.0$\times$, 1.8$\times$, and 1.2$\times$ for sequence lengths of 128K, 256K, 512K, and 1M, respectively. With Selective Checkpoint++, \SysName further boosts the training performance by 1.15$\times$, 1.18$\times$, 1.22$\times$, and 1.24$\times$ for the same sequence lengths. Consequently, Figure~\ref{fig:e2e_compare_64GPU}(a) shows that \SysName's overall training performance is improved by 5.2$\times$, 3.6$\times$, 2.3$\times$, and 1.5$\times$, respectively.
Additionally, we observe that to achieve higher training performance for MHA, \SysName tends to use a higher head parallelism size for sequence lengths of 128K and 256K. For sequence lengths of 512K and 1M, \SysName tends to use a balanced head and context parallelism size.

2D-Attention also works effectively for GQA. Compared to the performance of Ring-Attention, \SysName enhances the MFU for  sequences of 128K, 256K, 512K, and 1M by 1.58$\times$, 1.27$\times$, 1.11$\times$, and 1.07$\times$, respectively. Incorporating Selective Checkpoint++, \SysName further elevates the training performance by 1.21$\times$, 1.22$\times$, 1.23$\times$, and 1.25$\times$ for the same sequence lengths. Consequently, Figure~\ref{fig:e2e_compare_64GPU}(b) shows that the overall training performance is improved by 1.9$\times$, 1.5$\times$, 1.3$\times$, and 1.3$\times$, respectively. For GQA, a balanced head and context parallelism size is a more efficient configuration.

\subsection{Analysis of 2D-Attention}
\label{sec:exp_2d}

We evaluated 2D-Attention by measuring the average overall execution time and \texttt{SeqAlltoAll} communication time for a single 2D-Attention forward operation under various configurations. The results are presented in Table~\ref{tab:overall_a2a_time}. 
% Due to the constraints on the number of attention heads, we do not show the results for $d_{cp} = 1$.

\renewcommand{\arraystretch}{1}
\begin{table*}[htbp]
\centering
\small
\begin{tabular}{@{}c@{}|*{4}{>{\raggedleft\arraybackslash}p{0.6cm}}|*{4}{>{\raggedleft\arraybackslash}p{0.6cm}}|*{4}{>{\raggedleft\arraybackslash}p{0.6cm}}|*{4}{>{\raggedleft\arraybackslash}p{0.6cm}}@{}}
\toprule
 &  \multicolumn{4}{c|}{MHA (CP=64, HP=1)} & \multicolumn{4}{c|}{MHA (CP=16, HP=4)} & \multicolumn{4}{c|}{GQA (CP=64, HP=1)} & \multicolumn{4}{c}{GQA (CP=16, HP=4)} \\ \midrule
  Inner Ring Size  & 128K  & 256K  & 512K  & 1M & 128K  & 256K  & 512K  & 1M & 128K  & 256K  & 512K  & 1M & 128K  & 256K  & 512K  & 1M \\ \midrule
1 & 295.9 & 597.7 & 1214 & 2913 & 86.3 & 213.8 & 697.9 & 2621 & 94.2 & 226.7 & 713.5 & 2668 & 60.7 & 180.6 & 673.3 & 2567 \\
2 & 184.5 & 401.3 & 917.1  & 2823 & 72.6 & 205.7 & 710.7 & 2611 & 83.2 & 218.9 & 730.5 & \cellcolor{gray!25} 2650 & 60.8 & 182.6 & \cellcolor{gray!25} 671.2 & \cellcolor{gray!25} 2530 \\
4 & \cellcolor{gray!25} 140.6 & \cellcolor{gray!25} 316.3 & \cellcolor{gray!25} 842.7  & \cellcolor{gray!25} 2754 & \cellcolor{gray!25} 69.1 & \cellcolor{gray!25} 199.4 & \cellcolor{gray!25} 704.4 & \cellcolor{gray!25} 2610 & \cellcolor{gray!25} 78.4 & \cellcolor{gray!25} 210.3 & \cellcolor{gray!25} 719.7 & 2669 & \cellcolor{gray!25} 60.3 & \cellcolor{gray!25} 182.0 & 675.2 & 2535 \\
8 & 214.9 & 415.1 & 869.9  & 2815 & 77.4 & 198.7 & 705.3 & 2621 & 83.4 & 211.6 & 723.1 & 2674 & 61.0 & 183.1 & 677.4 & 2537 \\
\bottomrule
\end{tabular}
\caption{Average execution time (ms) of a single 2D-Attention  forward and backward operation (with Double-Ring-Attention and context-first device placement) on 64 GPUs with $d_{sp}=64$. The lowest execution time in each column is highlighted.}
\label{tab:doublering}
\end{table*}
\renewcommand{\arraystretch}{1}

\noindent \textbf{Sequence Length Study.}
As discussed in Section~\ref{sec:design_model}, 
% the computational complexity of attention is \(O(S^2D)\). When scaling the sequence length from 128K to 1M, the computational workload increases by  64$\times$, while the communication volume per \texttt{P2P} or \texttt{SeqAlltoAll} operation increases by  8$\times$. Therefore, 
with a fixed sequence parallelism degree, a longer sequence length provides more opportunities for computation-communication overlap. 
% As shown in Table~\ref{tab:overall_a2a_time}, 
When $d_{hp}=1$ and the sequence length grows from 128K to 1M, the overall attention time for MHA only increases by $9.7\times$, from 296.4ms to 2897ms, despite the computational workload increasing by $64\times$. In this configuration, there are no \texttt{SeqAlltoAll} operations, indicating that the primary performance bottleneck lies in \texttt{P2P} operations. In the case of GQA, the overall attention time increases from 86.0ms to 2681ms. Across all sequence lengths, GQA demonstrates a shorter execution time compared to MHA due to the reduced communication volume.

\noindent\textbf{MHA Study.} The execution time of MHA can be reduced significantly under the most appropriate configuration from Table~\ref{tab:overall_a2a_time}. Specifically, the execution time decreases from 296.4ms to 58.4ms when \SysName increases the head parallelism degree to 16 for 128K sequence length. When processing a sequence length of 1M, the overall execution time decreases from 2681ms to 2555ms when \SysName increases the head parallelism degree to 8. As discussed in Section \ref{sec:design_model}, the communication volume per \texttt{P2P} operation remains unaffected by \(d_{hp}\)(as long as $d_{sp}$ keeps the same), but the computation time per micro-step increases linearly with increased \(d_{hp}\). Therefore, \SysName can more effectively overlap the \texttt{P2P} communication with computation by increasing \(d_{hp}\), even though such a configuration introduces more \texttt{SeqAlltoAll} communication time.

\noindent\textbf{GQA Study.} GQA introduces less communication volume and is less sensitive to \(d_{cp}\) compared to MHA. For instance, processing a 128K sequence with \(d_{cp}=64\) results in an execution time of 86.0ms per GQA operation, which is 3.4$\times$ shorter than that of MHA. \SysName further reduces the GQA execution time by increasing \(d_{hp}\), thereby enhancing the ability to overlap \texttt{P2P} communication with computation. By increasing \(d_{hp}\) to 8, \SysName decreases the GQA execution time from 86.0ms to 56.8ms for a sequence length of 128K, and from 2681ms to 2495ms for a sequence length of 1M.
However, increasing \(d_{hp}\) beyond 8 does not further reduce the GQA execution time due to the significant increase in the \texttt{SeqAlltoAll} communication time. For example, when \(d_{hp}\) is increased from 8 to 32, the \texttt{SeqAlltoAll} communication time for processing a 128K sequence with head-first placement rises from 2.64ms to 13.40ms. 
% This is due to two factors: 1) the NCCL \texttt{AlltoAll} communication begins to cross nodes, and 2) the communication size increases due to KV replication.
In summary, to process GQA efficiently, the configuration of \(d_{hp} = 8\) and \(d_{cp} = 8\) avoids the large \texttt{SeqAlltoAll} overhead and effectively overlaps the computation with \texttt{P2P} communication.

\noindent\textbf{Device Placement Study.} 
As analyzed in Section~\ref{sec:design_model}, there is a trade-off between the \texttt{SeqAlltoAll} time and total execution time when choosing the placement strategy. Table~\ref{tab:overall_a2a_time} shows that when \(d_{cp}\) is large (e.g., \(d_{cp}=32\)), a single Attention operation can benefit from context-first placement. Although the context-first strategy increases the \texttt{SeqAlltoAll} time, the overall time is more advantageous due to the reduced \texttt{P2P} communication time. However, as \(d_{hp}\) gets larger, head-first placement performs better. In these cases, the increased large \texttt{SeqAlltoAll} volumes become the bottleneck of the overall execution time. Therefore, only if \texttt{SeqAlltoAll} leverages the intra-node high-bandwidth NVLINK can \SysName achieve better overall performance.

%\noindent \textbf{Double-Ring-Attention}

\noindent\textbf{Double-Ring-Attention Study. }
We compare the execution time of 2D-Attention with different inner ring sizes in Table~\ref{tab:doublering}. As expected, with MHA and shorter sequence length, \texttt{P2P} communication cannot be effectively overlapped with the computation. In these cases, Double-Ring-Attention achieves more speedup. For instance, when the sequence length is 128K and $d_{cp}=16$, Double-Ring-Attention further reduces the attention operation time by a factor of 1.2, even if 2D-Attention is already applied. However, with longer sequence lengths, due to the increased computational workload, the \texttt{P2P} communication can be overlapped more, limiting the improvements from Double-Ring-Attention. 

As we theoretically analyzed in Section~\ref{sec:design_model}, when the inner ring size matches the number of NICs in one node (4 in our case), all NICs can be utilized for outer-ring communication, which is more effective. Table~\ref{tab:doublering} also illustrates this trend. 
%We also observe that Double-Ring-Attention becomes more efficient when the \texttt{P2P} communication cannot be effectively overlapped with the computation. For instance, when the sequence length is 128K and $d_{cp}=16$, Double-Ring-Attention further reduces the attention operation time by a factor of 1.2, even if 2D-Attention is already applied. However, with longer sequence lengths, due to the increased computational workload, the \texttt{P2P} communication can be overlapped more, limiting the improvements from Double-Ring-Attention. 
As discussed, the global batch size poses a challenge for the computation-communication ratio when scaling $d_{sp}$ to 512 GPUs for a 1M sequence length. In such cases, Double-Ring-Attention is expected to be more useful.

%% file: 6_Conclusion.tex
\section{Conclusion}
We proposed \SysName, an efficient training framework for LLMs with long sequences. We designed the 2D-Attention, which combined both head-parallel and context-parallel approaches, to break the scalability constraints while maintaining high efficiency. We introduced the Double-Ring-Attention and device placement strategy to further improve the training efficiency. 
We implemented the \SysName system with hybrid parallelism and advanced gradient checkpoint techniques.
Experiment results showed that \SysName provides a significant performance improvement over existing systems, such as DeepSpeed-Ulysses and Megatron CP.

% To conclude, \SysName leverages 2D-Attention to optimize long sequence training, providing significant performance improvement over existing systems, such as DeepSpeed-Ulysses and Megatron Context Parallelism.  It uses the Double-Ring-Attention and device placement strategy to further improve the efficiency of  long sequence training. 
% We implemented the \SysName system with hybrid parallelism and advanced gradient checkpoint techniques.

\section{Acknowledgements}

We express our  gratitude to Zilin Zhu from Tencent. Our research benefited from his GitHub repository "ring-flash-attention," which implements Ring-Attention with FlashAttention. Additionally, we are thankful to Jiarui Fang and Shangchun Zhao from Tencent for their pioneering work in integrating Ulysses and Ring-Attention, as demonstrated in the open-source project Yunchang~\cite{fang2024unified}. Their guidance was instrumental in shaping this work. We also extend our thanks to Haoyu Yang and Jidong Zhai from Tsinghua University for their assistance in enhancing the performance of our implementation.

%% file: 7_Appendix.tex
\section{Appendix}

Table \ref{tab:e2e_mha_gqa_tgs} shows training performance (TGS) of 7B-MHA and 7B-GQA on 64 GPUs with $d_{sp}=64$.

\renewcommand{\arraystretch}{1}
\begin{table*}[]
\centering
\begin{tabular}{@{}p{0.05cm}p{0.3cm}p{0.3cm}|*{4}{>{\raggedleft\arraybackslash}p{0.65cm}}|*{4}{>{\raggedleft\arraybackslash}p{0.65cm}}|*{4}{>{\raggedleft\arraybackslash}p{0.65cm}}|*{4}{>{\raggedleft\arraybackslash}p{0.65cm}}@{}}
\toprule
& &    & \multicolumn{4}{c|}{128K}   & \multicolumn{4}{c|}{256K}  & \multicolumn{4}{c|}{512K}   & \multicolumn{4}{c}{1M}   \\
% \cmidrule(lr){4-7} \cmidrule(lr){8-11} \cmidrule(lr){12-15} \cmidrule(lr){16-19}
\midrule
& &    & \multicolumn{2}{c}{With SC++} & \multicolumn{2}{c|}{W/O SC++} & \multicolumn{2}{c}{With SC++} & \multicolumn{2}{c|}{W/O SC++} & \multicolumn{2}{c}{With SC++} & \multicolumn{2}{c|}{W/O SC++} & \multicolumn{2}{c}{With SC++} & \multicolumn{2}{c}{W/O SC++} \\
% \cmidrule(lr){4-7} \cmidrule(lr){8-11} \cmidrule(lr){12-15} \cmidrule(lr){16-19}
\midrule
& $d_{cp}$ & $d_{hp}$ & HF  & CF  & HF & CF & HF   & CF   & HF  & CF  & HF   & CF   & HF & CF  & HF   & CF  & HF & CF           \\
\midrule
\multirow{6}{*}{\rotatebox{90}{MHA}} 
& 64 & 1  & 190.2 & 190.2 & 145.3 & 145.3 & 195.4 & 195.4 & 149.4 & 149.4 & 196.8 & 196.8 & 149.9 & 149.9 & 161.7 & 161.7 & 127.6 & 127.6 \\
& 32 & 2  & 203.9 & 327.1 & 158.8 & 260.4 & 212.0 & 340.8 & 163.6 & 269.2 & 214.2 & 294.3 & 164.7 & 239.3 & 169.8 & 173.9 & 140.8 & 145.2 \\
& 16 & 4  & 363.2 & 505.9 & 290.4 & 422.5 & 386.0 & 464.6 & 304.7 & 389.1 & 318.7 & 319.7 & 260.0 & 262.7 & 185.7 & 182.1 & 149.3 & 147.5 \\
& 8  & 8  & 585.6 & 662.6 & 486.9 & 582.2 & 533.5 & 515.6 & 443.6 & 437.8 & \cellcolor{gray!25}  340.1 & \cellcolor{gray!25}  324.0 & \cellcolor{gray!25}  277.1 & \cellcolor{gray!25}  \cellcolor{gray!25} 266.8 & \cellcolor{gray!25} 188.4 & \cellcolor{gray!25} 186.1 & \cellcolor{gray!25} 151.7 & \cellcolor{gray!25} 150.2 \\
& 4  & 16 & \cellcolor{gray!25}  676.9 & 675.9 & \cellcolor{gray!25}  596.3 & 585.0 & \cellcolor{gray!25}  535.2 & \cellcolor{gray!25}  519.5 & \cellcolor{gray!25}  452.4 & \cellcolor{gray!25}  441.1 & 329.9 & 323.0 & 270.4 & \cellcolor{gray!25} 266.8 & 185.5 & 185.9 & 149.3 & 147.2 \\
& 2  & 32 & 661.0 & \cellcolor{gray!25}  679.9 & 
 586.7 & \cellcolor{gray!25}  605.7 & 516.4 & 517.2 & 433.6 & 438.7 & 321.3 & 323.8 & 263.2 & 267.2 & 185.0 & 185.0 & 148.4 & 145.0 \\

\hline
\multirow{6}{*}{\rotatebox{90}{GQA}} & 64 & 1  & 526.0 & 526.0 & 404.8 & 404.8& 465.4 & 465.4 & 377.6 & 377.6 & 318.7 & 318.7 & 256.5 & 256.5 & 181.6 & 181.6 & 145.3 & 145.3 \\
& 32 & 2  & 585.3 & 655.0 & 480.6 & 555.4 & 514.6 & 527.2 & 424.0 & 435.1 & 333.5 & 328.5 & 270.0 & 265.9 & 186.4 & 184.6 & 149.5 & 148.9 \\
& 16 & 4  & 732.1 & 698.8 & 637.6 & 606.6 & 571.6 & 537.0 & 473.1 & 457.6 & 342.4 & 334.8 & 277.7 & 273.6 & 189.7 & 187.9 & 152.1 & 152.4 \\
& 8  & 8  & \cellcolor{gray!25} 779.7 & \cellcolor{gray!25} 730.6 & \cellcolor{gray!25}  676.0  & \cellcolor{gray!25} 640.8 & \cellcolor{gray!25} 588.9 &\cellcolor{gray!25}  554.7 & \cellcolor{gray!25} 481.3 & \cellcolor{gray!25} 466.4 & \cellcolor{gray!25} 349.8 & \cellcolor{gray!25} 340.6 & \cellcolor{gray!25} 284.3 & \cellcolor{gray!25} 279.2 & \cellcolor{gray!25} 194.0 & \cellcolor{gray!25} 191.6 & \cellcolor{gray!25} 155.6 & \cellcolor{gray!25} 154.3 \\
& 4  & 16 & 731.2 & 705.1 & 641.0 & 636.5 & 561.1 & 536.1 & 463.1 & 458.5 & 339.1 & 334.2 & 277.0 & 274.3 & 190.1 & 189.2 & 152.9 & 149.8 \\
& 2  & 32 & 666.4 & 687.5 & 589.2 & 609.7 & 520.3 & 517.6 & 428.1 & 441.6 & 322.8 & 325.9 & 264.0 & 267.3 & 185.1 & 185.1 & 148.3 & 145.0 \\

\bottomrule
\end{tabular}
\caption{End-to-End Training Performance (TGS) of 7B-MHA and 7B-GQA on 64 GPUs with $d_{sp}=64$. SC++ stands for \textit{Selective-Checkpoint++}, HF for head-first, and CF for context-first. The highest TGS value in each column is highlighted.}
\label{tab:e2e_mha_gqa_tgs}
\end{table*}
\renewcommand{\arraystretch}{1}